\documentclass[aps,prd,reprint,a4paper,showpacs,nofootinbib,superscriptaddress]{revtex4-2}

\usepackage{hyperref}
\hypersetup{
  colorlinks=true,        
  linkcolor=blue,         
  citecolor=cyan,         
}

\providecommand{\dif}{\mathrm{d}} \def\d{\dif}
\usepackage{graphicx}
\usepackage{dcolumn}
\usepackage{bm}
\usepackage{color}
\usepackage{enumitem}
\usepackage{amsmath}
\providecommand{\bibfnamefont}[1]{#1}
\providecommand{\bibnamefont}[1]{#1}
\providecommand{\bibinfo}[2]{#2}
\providecommand{\bibinfo}[1]{#1}

\usepackage{amssymb}
\def\EE{{\cal E}}
\def\LL{{\cal L}}
\newcommand{\beq}{\begin{equation}}
\newcommand{\eeq}{\end{equation}}
\newcommand{\bea}{\begin{eqnarray}}
\newcommand{\eea}{\end{eqnarray}}
\newcommand{\non}{\nonumber}
\usepackage{bbm}
\usepackage{amsfonts}
\usepackage{mathrsfs}
\usepackage{latexsym}
\usepackage{epsfig}
\usepackage{epstopdf}
\usepackage{epstopdf}
\usepackage{graphicx}
\usepackage{amssymb}
\usepackage{amsmath}
\usepackage{dcolumn}
\usepackage{bm}
\usepackage{color}
\usepackage{comment}
\usepackage{xcolor}
\usepackage{dsfont}

\usepackage{amsmath}
\begin{document}

\title{\bf Origin of Quasi-Periodic Oscillations and Accretion Process in X-Ray Binaries around  Quantum Lee-Wick Black Hole}


\author{Orhan~Donmez}
\email{orhan.donmez@aum.edu.kw}
\affiliation{College of Engineering and Technology, American University of the Middle East, Egaila 54200, Kuwait}

\author{G. Mustafa}
\email{gmustafa3828@gmail.com}
\affiliation{Department of Physics, Zhejiang Normal University, Jinhua 321004, China}

\author{M. Yousaf}
\email{myousaf.math@gmail.com}
\affiliation{Department of Mathematics, Virtual University of Pakistan, 54-Lawrence Road, Lahore 54000, Pakistan}

\author{Faisal Javed}
\email{faisaljaved.math@gmail.com}
\affiliation{College of Transportation, Tongji University, Shanghai 201804, People’s Republic of China}

\author{Ikhtiyor Saidov}
\email{ixti06.27@gmail.com}
\affiliation{Tashkent State Technical University, Tashkent 100095, Uzbekistan}

\author{Farruh Atamurotov}
\email{atamurotov@yahoo.com}
\affiliation{Urgench State University, Kh. Alimdjan str. 14, Urgench 220100, Uzbekistan}

\begin{abstract}
In this study, we investigate the accretion dynamics and test particle motion around a non-rotating, spherically symmetric Lee-Wick black hole (BH) to reveal how the model parameters affect orbital stability and the quasi-periodic oscillations (QPOs) observed in X-ray binary systems. The spacetime geometry, characterized by the BH mass and the coupling parameters $S_1$ and $S_2$, includes exponential and oscillatory corrections arising from the Lee-Wick terms. Using the effective potential approach, we derive specific energy, angular momentum, epicyclic frequencies, and the locations of the innermost stable circular orbits (ISCOs) of test particles. In addition to the analytical analysis, we explore the effects of the Lee-Wick spacetime parameters on the shock-cone morphology produced by Bondi-Hoyle-Lyttleton (BHL) accretion. To this end, we perform general relativistic hydrodynamic simulations in two characteristic regimes: Block-1 (weak Lee-Wick regime) and Block-2 (strong Lee-Wick regime). The results show that Block-1 solutions closely resemble the Schwarzschild case, while Block-2 models develop denser and asymmetric shock cones accompanied by stronger QPOs activity, shifting from low-frequency to high-frequency QPOs. These variations yield distinct observational signatures that may be detectable in high-resolution X-ray timing data. Our analytical and numerical findings demonstrate that the Lee-Wick parameters $S_1$ and $S_2$ cause measurable changes in the morphology of the accretion flow and in the frequency ratios near the BH. This suggests that future multi-wavelength observations could provide an important avenue to test higher-derivative gravity theories.
\end{abstract}

\maketitle

\date{\today}

\section{Introduction}

In recent years, there has been a lot of interest in the perturbative approach to quantum gravity, especially because of the new understanding of the difficulties presented by higher derivatives and related ghost fields \cite{Shapiro:2022ugk}. Several models and processes have been proposed to reconcile quantum gravity's unitarity with renormalizability. Interestingly, models of higher-derivative gravity with complicated poles in the propagator known as Lee-Wick gravity have been investigated~\cite{Modesto:2015ozb, Modesto:2016ofr}. The theory becomes super-renormalizable when larger derivatives are added to the action, improving the propagator's behavior in the ultraviolet limit~\cite{Asorey:1996hz}. Similarly to the initial concepts proposed by Lee and Wick~\cite{Lee:1969fy, Lee:1970iw}, the ghost-like poles that occur in complex conjugate pairs contribute to a unitary $S$-matrix ~\cite{Cutkosky:1969fq, Anselmi:2017lia, Liu:2022gun}.

Recent advances in understanding the problems of higher derivatives and the corresponding ghost fields have led to renewed interest in the perturbative approach to quantum gravity. To reconcile renormalizability and unitarity, two essential components of quantum gravity researchers have put forth several models and processes. One prominent method is the creation of models of higher derivative gravity, also known as Lee-Wick gravity~\cite{Modesto:2015ozb, Modesto:2016ofr}, which incorporate complex poles in the propagator. In other words, the propagator behaves better at high energies (also known as the ultraviolet regime) when greater derivatives are added to the action. A super-renormalizable theory~\cite{Asorey:1996hz} is the result of this enhancement. Moreover, ghost-like poles of complex conjugate pairs contribute to the maintenance of a unitary $S$ matrix, guaranteeing that probabilities remain constant throughout time~\cite{Modesto:2015ozb}. According to the relevant literature~\cite{Cutkosky:1969fq, Anselmi:2017lia, Liu:2022gun}, this idea is reminiscent of the initial suggestions made by Lee and Wick in the studies they conducted on quantum field theory~\cite{Lee:1969fy, Lee:1970iw}.

One of the most prominent features of complex poles at the classical level is the oscillating behavior seen in linearized solutions to the equations of motion~\cite{Modesto:2016ofr, Accioly:2016qeb, Giacchini:2016xns}. Both theoretical and experimental investigations have investigated these oscillations, which contribute to gravitational force, especially in the low-energy regime~\cite{Accioly:2016etf, Boos:2018bhd, Perivolaropoulos:2016ucs, Antoniou:2017mhs, Krishak:2020opb}. However, higher derivatives make the analysis more challenging and more complicated in obtaining precise solutions to the entire set of field equations. To address this complexity, scientists have created an empirical representation of Lee-Wick BHs that takes these oscillatory solutions into account~\cite{Bambi2017PLB}. The purpose of this model is to have a better understanding of how the oscillations affect BH behavior in the context of Lee-Wick gravity. 

Recent research has investigated several important features of Lee-Wick BHs associated with the smeared source. The regularity of curvature and the thermodynamic properties of BHs~\cite{BreTib2,Bambi2017PLB,Giacchini2023,Burzilla2023JCAP, dePaulaNetto:2023vtg}, the characteristics and conditions of the horizons for various sizes of BH~\cite{Burzilla2023JCAP}, the BHs that are impacted by noncommutative geometry~\cite{Nicolini:2005vd, Nicolini:2009gw}, gravitational light deflection~\cite{Zhao:2017jmv, Buoninfante:2020qud, Zhu:2020wtp}, and the precession of orbits~\cite{Lin:2022wda} are just a few of the topics covered in these investigations. The Newman-Janis algorithm~\cite{Singh:2022tlo} has also been used to work on a spinning extension of these BHs. An especially intriguing discovery, as stated in~\cite{Burzilla2023JCAP}, is that the ratio \(q = b/a\), where \(b\) and \(a\) reflect particular characteristics associated with the BH structure, affects the oscillation pattern of solutions. This ratio determines acceptable limits for the placement of the event horizon in addition to influencing the possible number of horizons. In the end, the ratio \(q\) and the mass \(M\) of the effective source define the true number of BH horizons. A deeper understanding of the characteristics and behavior of Lee-Wick BHs under various physical circumstances may result from this knowledge. Batic et al. \cite{paper} used the spectrum approach to investigate scalar, electromagnetic, and gravitational perturbations in their thorough analysis of quasinormal modes associated with Lee-Wick BHs.  Our research takes into account all representative cases, both extremal and non-extremal, and shows that such BHs can exhibit a rich structure of horizons.  Specifically, they demonstrated that for extreme and near-extremal configurations, entirely imaginary quasinormal modes appear, indicating a quick return to equilibrium without oscillation.

The sixth-derivative gravity field equations cannot be solved exactly by the Lee-Wick BHs suggested in~\cite{Bambi2017PLB, Burzilla2023JCAP}, but they might capture some of their key characteristics, namely curvature regularity and the existence of many or extreme horizons. The field equations generated from the extended Einstein-Hilbert action, containing all possible higher-derivative terms with up to six metric derivatives, have recently been subjected to a thorough investigation by researchers. They discovered that a bounded Kretschmann scalar \(R_{\mu\nu\alpha\beta}^2\)~\cite{Giacchini:2024exc} is present in all vacuum solutions that allow Frobenius expansion. 
Its equivalent to the Schwarzschild solution in general relativity appears to be a regular metric~\cite{Giacchini:2024exc}. The current research and knowledge is limited to the weak field regime or deals with local features of the solutions~\cite{Holdom:2002xy, Giacchini:2024exc}. Due to this constraint, it is currently impossible to say with certainty if the horizon structures described in~\cite{Burzilla2023JCAP} match those of the precise solutions. However, it is interesting to note that the exact solutions of sixth-derivative gravity~\cite{Giacchini:2024exc} contain both simple BHs and extreme horizons, confirming the requirement that a regular, asymptotically flat geometry have an even number of horizons~\cite{Holdom:2002xy}.

QPOs are rapid fluctuations in $X$-ray emissions from various astrophysical sources, such as Active Galactic Nuclei (AGN), high-mass X-ray binaries (HMXBs), low-mass X-ray binaries (LMXBs) and ultraluminous X-ray systems \cite{Pasham2018ALQ, Smith_2021, Singh2022MNRAS}. In LMXBs and HMXBs, BHs at their centers constantly accrete matter from a partner star \cite{charles2003optical, Walter_2015}. These oscillations are considered to be caused by the intricate interactions of matter as it spirals inward into the BH. QPOs are formed as a result of these interactions, with oscillation frequencies ranging from milliHertz to several hundred Hertz \cite{fragile_2016}. The QPOs in AGNs can be discovered using measurements throughout the electromagnetic spectrum, including radio waves and gamma rays \cite{Ackermann2015ApJ, Zhang_2021}. However, it is crucial to note that this phenomenon is more commonly found in systems with stellar-mass BHs, as evidenced by X-ray spectrum data analysis.

The study of BHs has gone beyond the classical scope of general relativity to include aspects of quantum gravity, dark matter phenomenology, and horizon-scale studies. The regular BH models which do not admit curvature singularities have garnered enthusiasm for the reconciliation of the quantum spacetime geometry. Several works have shown how non-minimal couplings of scalars and the curvature will change Hawking BH radiation to modify the thermodynamics of BHs, such as those proposed by Garfinkle \cite{v1}. Also, primordial regular BHs emerging from the effective quantum-gravity theories have been suggested as candidates for black matter, both for tr-symmetric and non-symmetric spacetime structures \cite{v2,v3,v4}. The dynamics of BHs and the associated gravitational-wave signatures it emits have also been studied, particularly those signature of dark-energy cosmological evolution \cite{v5}. Such ideas have encouraged the study of the interaction of BHs with large-scale structures. The many revolutionary results from the Event Horizon Telescope have provided new avenues for the study of BHs and their shadows to test fundamental theories of physics.
Horizon-scale imaging has been used to rule out black-hole solutions in mimetic gravity \cite{v6}, carry out the most detailed constraints on the beyond-GR effects around Sgr A, and investigate the possible presence of ultralight bosonic particles via shadow distortions and superradiant instabilities. In addition, studies on rotating regular BHs have uncovered significant connections between quasi-normal modes and shadow features \cite{v7}. Other works have investigated the shadows of BHs with scalar hair \cite{v8}, their role as standard candles in cosmology \cite{v9}, and the shadow resulting from the presence of extra dimensions \cite{v10} or possible overspinning states of M87 \cite{v11}.

Low-frequency quasi-periodic oscillations (LFQPOs) discovered from different sources have a frequency range of a few millihertz to many tens of Hertz. To understand these frequencies, it may be necessary to look at the physical structures formed by the matter accreting around BHs. This will help reveal important characteristics like the mass and angular momentum of the BHs in the cores of these systems. Three different forms of LFQPOs have been identified by observations from sources such as XTE J1550-564: types A, B, and C \cite{Homan2001ApJS, belloni2011black, Belloni2024MNRAS}. Type-A QPOs have been found in about ten distinct sources and are quite uncommon in BH binary systems. They usually appear during the transition from the hard state to the soft state. Type-B QPOs are mostly associated with the physical processes behind astrophysical jets and have been seen in a small number of BH binary systems. The X-ray spectrum studies are usually used to detect Type-C QPOs, which are the most common in these systems \cite{Motta_2012, Motta2016AN, Liu_2021}. GRS 1915+105 is an excellent example of a Type-C QPO. It has been the focus of numerous theoretical and numerical investigations to clarify the physical properties of BH and the phenomena that take place around it \cite{Belloni2024MNRAS, Belloni2013MNRAS, Misra2020ApJ, wang20242022, Chauhan2024MNRAS}. The dynamic changes in the matter flow around the BH's event horizon are thought to be the cause of these oscillations. 

The orbital dynamics of test particles in the framework of non-rotating Lee-Wick BH is the main topic of this publication. An outline of the structure of the paper may be seen in the following:
We present a comprehensive review of Lee-Wick BH in Section II. The ISCOs, equatorial circular orbits are discussed in the next section. We use an effective potential framework to examine the stability of these orbits. The oscillating motion of test particles around the Lee-Wick BH is examined in Section IV, which also covers the phenomenon of periastron precession. In addition, we present the dynamic structure of the shock cone produced by BHL accretion around a Lee-Wick BH by using numerical simulation in Section V. The observation constraints of accretion regimes are presented in Section VI. In the last section, we present the concluding remarks. 

\section{Lee-Wick Gravity and Black Holes} \label{BH}

\begin{figure*}
\centering 
\includegraphics[width=\hsize]{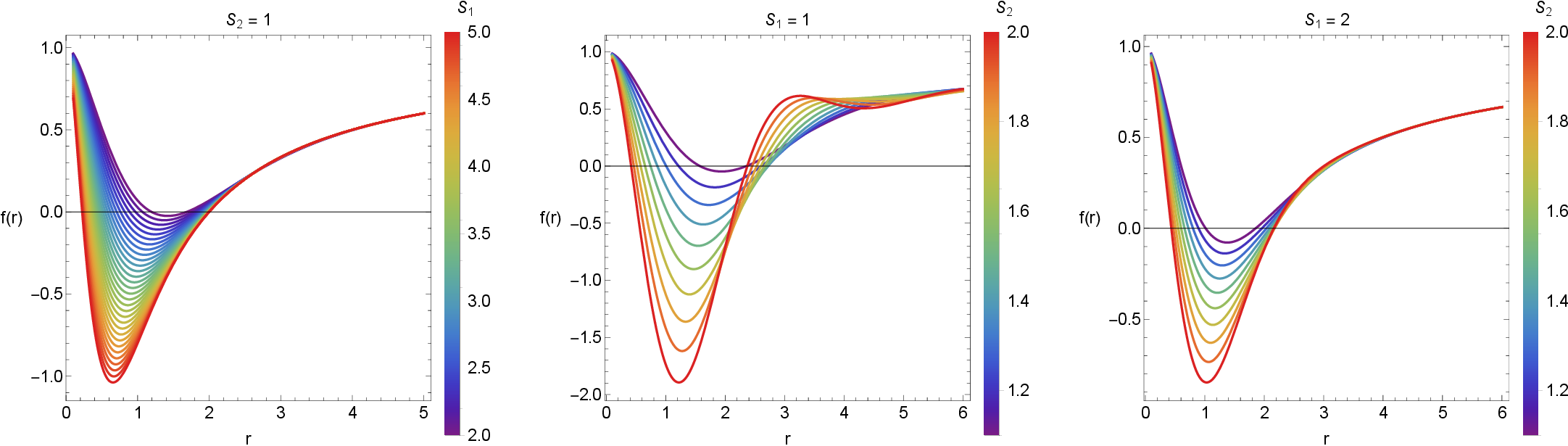}
\caption{Behaviour of lapse function of the spacetime of non-rotating Lee-Wick BH. 
}\label{fig_LapseFun}
\end{figure*}

The effective field equations can be expressed as \cite{Bambi2017PLB}
\begin{equation}
\label{Gff}
G^{\mu} {}_{\nu} = 8 \pi G \, \tilde{T}^{\mu} {}_{\nu},
\end{equation}
where the Newton constant is denoted by $G$ and the Einstein tensor is represented by $G_{\mu\nu}$. The effective stress-energy tensor becomes
\begin{equation}
\label{effT}
\tilde{T}^{\mu} {}_\nu = \text{diag}(-\rho , p_r , p_\theta , p_\theta ).
\end{equation}
It would replicate the effect of the higher derivatives~\cite{Bambi2017PLB}. For a sixth-derivative Lee-Wick gravity model that is determined by the action
\begin{equation}
\label{action}
S = \frac{1}{16 \pi G} \int d^4 x \sqrt{-g} \left[ R + G_{\mu\nu} \left( S_1 + S_2 \Box \right)  R^{\mu\nu} \right], 
\end{equation}
with the conditions $S_2> 0,\;-2\sqrt{S_2}< S_1<2 \sqrt{S_2}$ and
the respective source leads to ~\cite{BreTib2,Bambi2017PLB,Giacchini2023}
\begin{equation}\label{effsource}
 \rho(r)  =  \frac{M (a^2+b^2)^2}{8 \pi ab } \frac{e^{-ar}\sin(b r) }{r},
\end{equation}
here, the total source mass is denoted by $M$.
With inverse length dimensions, the parameters $a$ and $b$ are connected to the couplings at work through
\begin{equation}\label{Rel2}a^2 = \frac{2 \sqrt{S_2} - S_1}{4 S_2}, \qquad
  b^2 = \frac{2 \sqrt{S_2} + S_1}{4 S_2} .
\end{equation}
The real and imaginary parts of the Lee-Wick mass $m = a + i b$ are represented by them, respectively. Note that $a,b\neq 0$ and $a,b\in R$ are guaranteed by the domain of the parameters $S_1$ and $S_2$ in~\eqref{action}. By adding an effective equation of state and making sure that the continuity equation \(\nabla_\mu \tilde{T}^\mu {}_\nu = 0\) is satisfied~\cite{Burzilla2023JCAP, dePaulaNetto:2023vtg, Bambi2017PLB}, the remaining components of \(\tilde{T}\) in Eq. (\ref{effT}) can be identified. The method used for BHs that are affected by noncommutative geometry~\cite{Nicolini:2005vd, Nicolini:2009gw} and other effective metrics~\cite{Giacchini2023} is comparable to this. The line element describing the geometry of spherically symmetric Lee-Wick BH is given by \cite{paper}
\beq
    \d s^2 = -f(r) \d t^2 + f^{-1}(r) \d r^2 + r^2(\d \theta^2 + \sin^2\theta \d \phi^2), \label{metric}
\eeq
where the lapse function $f(r)$ reads
\begin{widetext}
\begin{eqnarray}
    f(r)= 1-\frac{2 M}{r}\left[1-\frac{\exp (-r S_1)}{2 S_1 S_2} \left( \left(r \left(S_1^2+S_2^2\right) S_1+S_1^2-S_2^2\right) \sin \left(r S_2\right)+S_2 \left(r \left(S_1^2+S_2^2\right)+2 S_1\right) \cos \left(r S_2\right) \right) \right].
 \label{metric2}   
\end{eqnarray}
\end{widetext}
Here, $M$ is the mass of the gravitational source. The parameters $S_1$ and $S_2$ are related to the so-called Lee-Wick ``mass'' by $\tilde{m} =  S_1 + i S_2$ with $S_1, S_2 >0$. For this selection of the sign for the parameters $S_1$ and $S_2$, one can see from \cite{paper}. Horizons exist at those values of $r$ where $f(r)$ not only vanishes but also changes signs when crossing these points. Furthermore, the case of a Schwarzschild BH is recovered in the asymptotic regime $r>>1/S_1$. The behavior of the lapse function $f(r)$ as a function of $r$ for different values of the parameters $S_1$ and $S_2$ is shown in Fig. (\ref{fig_LapseFun}). The first column is plotted for setting $S_2 =1$, and varying the parameter $S_1$. The second and third columns are plotted to fix $S_1=1$ and $S_1=2$, respectively, while varying $S_2$. We see in all cases that the function $f(r)$ crosses the axis $f(r)=0$. This means that for these ranges of parameters, the BH solution exists. For parameter $S_1=1$, the lapse function changes as the parameter $S_2$ changes with increasing radius $r$. However, for parameter $S_1=2$, the lapse function does not change as parameter $S_2$ changes with increasing radius $r$. 

\section{Particle dynamics around Lee-Wick BH} \label{part_dynam}

The motion of a neutral particle can be described by the Hamiltonian given by
\beq
H=\frac{1}{2}g^{\mu \nu} p_{\mu} p_{\nu} + \frac{1}{2}\mu^2\label{Ham},
\eeq
where $\mu$ is the mass of the particle, $p^{\alpha}=\mu u^{\alpha}$ represents the four-momentum, $u^{\alpha}= d x^{\alpha}/d\tau$ denotes the four-velocity, and $ \tau $ is the proper time of the test particle. The Hamilton equations of motion can be written as
\beq
\frac{dx^{\alpha}}{d\zeta}\equiv \mu u^{\alpha}=\frac{\partial H}{\partial p_{\alpha}}, \quad
\frac{d p_{\alpha}}{d\zeta} = -\frac{\partial H}{\partial x^{\alpha}},
\eeq
here $ \zeta=\tau/\mu$ is the affine parameter. Due to the symmetries of the BH geometry, there exist two constants of motion, namely specific energy $E$ and specific angular momentum $L$, given by
\bea\label{EE}
\frac{p_{t}}{m}&=& g_{t t} u^t+g_{t \phi} u^\phi =-\mathcal{E}, \\\label{LL}
\frac{p_{\phi}}{m}&=& g_{\phi \phi} u^\phi+g_{t \phi} u^t =\mathcal{L},
\eea

where $\mathcal{E}=E/m$, $\mathcal{L}=L/m$ are the specific energy and specific angular momentum. The time $u^t$, azimuthal $u^\phi$, and radial $u^r$ components of the four-velocity $u^\alpha$ satisfy the following equations of motion
\begin{widetext}
\bea
\dot{t} &=& \frac{\mathcal{E}}{1-\frac{2 M}{r}\left[1-\frac{\exp (-r S_1)}{2 S_1 S_2} \left(\left(r \left(S_1^2+S_2^2\right) S_1+S_1^2-S_2^2\right) \sin \left(r S_2\right)+S_2 \left(r \left(S_1^2+S_2^2\right)+2 S_1\right) \cos \left(r S_2\right) \right)\right]},\\
\dot{\phi} &=& \frac{\mathcal{L}}{r^2 \sin^2\theta},\\\non
\dot{r}^2 &+& \left(\epsilon + \frac{\LL^2}{r^2 \sin^2\theta} \right)(1-\frac{2 M}{r} [1-\frac{\exp (-r S_1)}{2 S_1 S_2} (\left(r \left(S_1^2+S_2^2\right) S_1+S_1^2-S_2^2\right) \sin \left(r S_2\right) \\\ &+& S_2 \left(r \left(S_1^2+S_2^2\right)+2 S_1\right) \cos \left(r S_2\right) )] ) = \mathcal{E}^2,
\eea
\end{widetext}
where $\epsilon =1$ for timelike particles and $\epsilon =0$ for null-like particles. The dot denotes the derivative with respect to proper time $\tau$. In our work, we focus only on time-like particles. The Hamiltonian by Eq.(\ref{Ham}), for a non-rotating Lee-Wick BH can be written in the form
\begin{equation}\non
H =\frac{1}{2} \left(f(r) \right) p_r^2 +\frac{1}{2r^2} p_\theta^2  + \frac{1}{2} \frac{m^2}{f(r)} \left[ V_{\rm eff}(r,\theta) - \EE^2 \right],
\end{equation}
where the effective potential $V_{eff}(r, \theta)$ takes the form
\begin{equation}\non
V_{eff}(r, \theta)= \left(1 + \frac{\mathcal{L}^2 \csc^2 \theta }{r^2} \right) f(r).
\end{equation}

\subsection{Effective potential}

\begin{figure*}
\centering 
\includegraphics[width=\hsize]{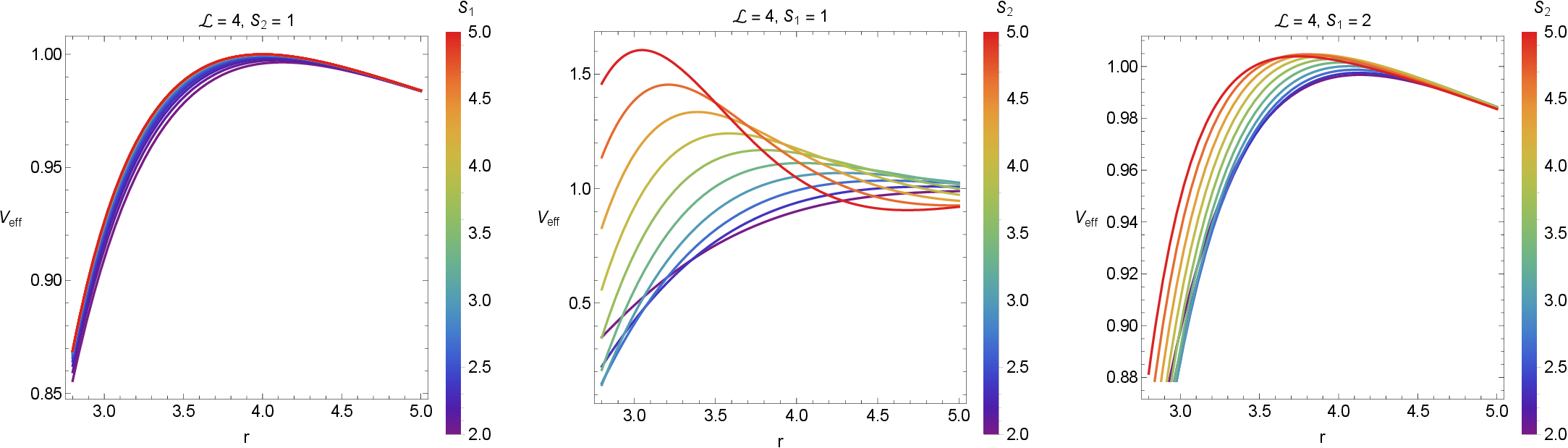}
\caption{Plots of the effective potential of particles moving around a non-rotating Lee-Wick BH.
}\label{fig_Veff}
\end{figure*}

The effective potential $V_{eff}(r, \theta)$ is crucial for understanding the motion of the test particles, as it allows one to describe the trajectories of the particles without directly solving the equations of motion. The maximum and minimum values of $V_{eff}$ correspond to unstable and stable circular orbits, respectively. Fig. (\ref{fig_Veff}) illustrates how $V_{eff}$ varies with $r$. Stable circular orbits are associated with the minima, whereas unstable orbits occur at the maxima of the potential. The circular orbits for the equatorial plane are found by the following conditions
\beq
V_{\rm eff}(r) = \EE^2, \quad \frac{\d V_{\rm eff} (r)}{\d r} = 0,
\label{Veff-1}
\eeq
to solve the Eq. (\ref{Veff-1}), we obtain the circular orbits around a non-rotating Lee-Wick BH 
\begin{widetext}
\beq\label{angMom}
\LL = \frac{r \sqrt{\frac{e^{-r S_1} \left(-\left(r^2 S_2^4+S_2^2 \left(r S_1+1\right) \left(2 r S_1-1\right)+r S_1^3 \left(r S_1+1\right)+S_1^2\right) \sin \left(r S_2\right)+2 S_1 S_2 e^{r S_1}-\left(S_2 \left(r S_2^2+S_1 \left(r S_1+2\right)\right) \cos \left(r S_2\right)\right)\right)}{S_1 S_2}}}{\sqrt{\frac{e^{-r S_1} \left(\left(r^2 S_2^4+S_2^2 \left(r S_1 \left(2 r S_1+3\right)-3\right)+S_1^2 \left(r S_1 \left(r S_1+3\right)+3\right)\right) \sin \left(r S_2\right)+2 (r-3) S_1 S_2 e^{r S_1}+3 S_2 \left(r S_2^2+S_1 \left(r S_1+2\right)\right) \cos \left(r S_2\right)\right)}{S_1 S_2}}},
\eeq
\end{widetext}
and the corresponding energy function takes the form 
\begin{widetext}
\begin{equation}\label{energy}
    \EE = \frac{\sqrt{2} e^{-r S_1} \left((r-2) S_1 S_2 e^{r S_1}+\left(r S_1^3+r S_2^2 S_1+S_1^2-S_2^2\right) \sin \left(r S_2\right)+S_2 \left(r S_1^2+r S_2^2+2 S_1\right) \cos \left(r S_2\right)\right)}{S_1 S_2 \sqrt{\frac{r e^{-r S_1} \left(\left(r^2 S_1^4+S_1^2 \left(2 r^2 S_2^2+3\right)+S_2^2 \left(r^2 S_2^2-3\right)+3 r S_1^3+3 r S_2^2 S_1\right) \sin \left(r S_2\right)+2 (r-3) S_1 S_2 e^{r S_1}+3 S_2 \left(r S_1^2+r S_2^2+2 S_1\right) \cos \left(r S_2\right)\right)}{S_1 S_2}}}.
\end{equation}
\end{widetext}

\begin{figure*}
\centering 
\includegraphics[width=\hsize]{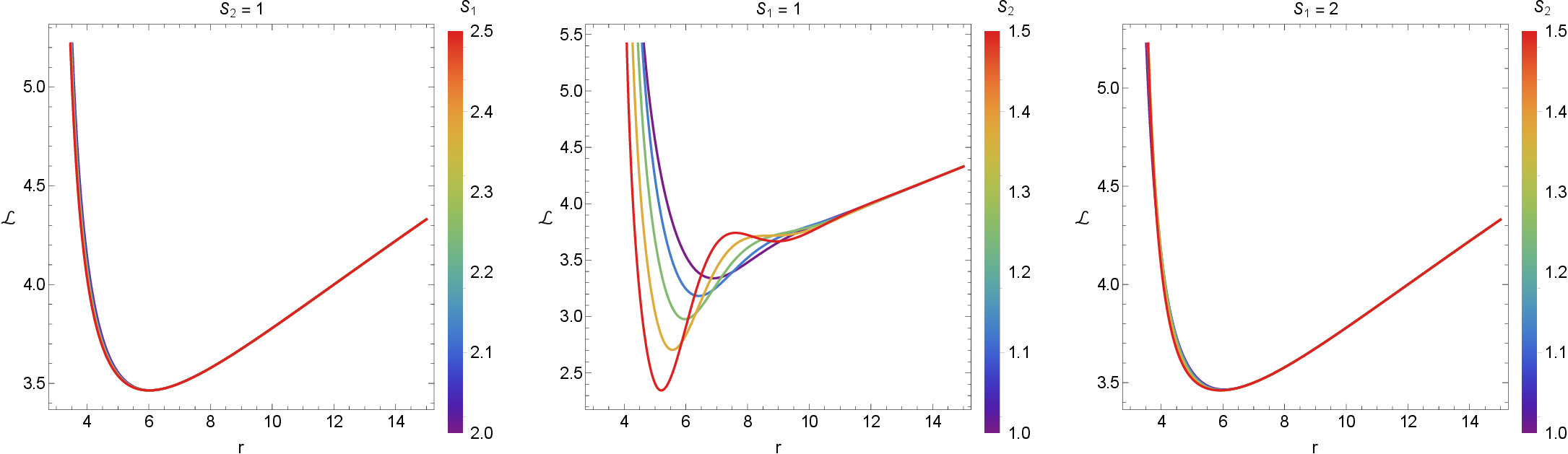}
\caption{Plots of angular momentum of particles moving non-rotating Lee-Wick BH.
}\label{fig_AngMom}
\end{figure*}

Fig. (\ref{fig_AngMom}) shows the angular momentum $\LL$ of equatorial circular orbits around a non-rotating Lee-Wick BH. The first column illustrates the influence of the parameter $S_1$ on the radial profiles of the angular momentum. We observe that there is not much difference in the angular momentum while varying the parameter $S_1$. The second and third columns are plotted for $S_1=1$ and $S_1=2$, respectively, varying the parameter $S_2$. For $S_1$, the angular decreases with increasing values of the parameter $S_2$. However, when $S_1=2$, the parameter $S_2$ does not change dramatically. 

\begin{figure*}
\centering 
\includegraphics[width=\hsize]{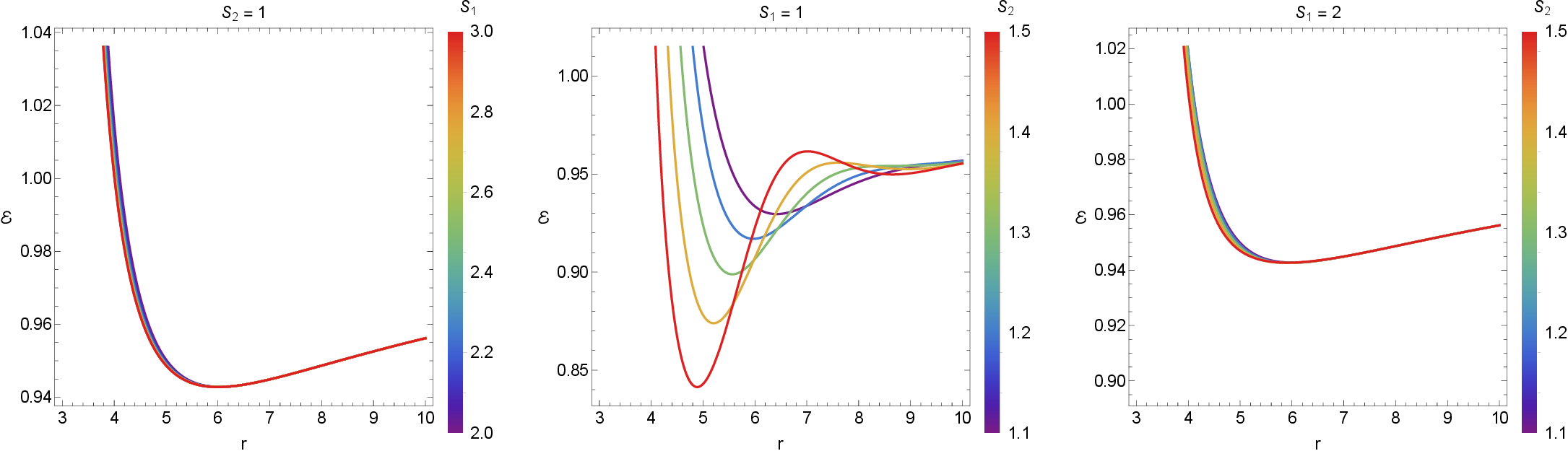}
\caption{Plots of energy of particles moving around a non-rotating Lee-Wick BH.
}\label{fig_Energy}
\end{figure*}

Fig. (\ref{fig_Energy}) presents the energy $\EE$ of equatorial circular orbits around a non-rotating Lee-Wick BH. The first panel shows how parameter $S_1$ affects the radial profiles of energy, while the second and third panels illustrate the impact of parameter $S_2$. As the parameter $S_1$ increases, the energy of the particles decreases. For $S_1$, an increase in the parameter $S_2$ results in a decrease in particle energy.


\subsection{Innermost stable circular orbits}

\begin{figure*}
\centering 
\includegraphics[width=\hsize]{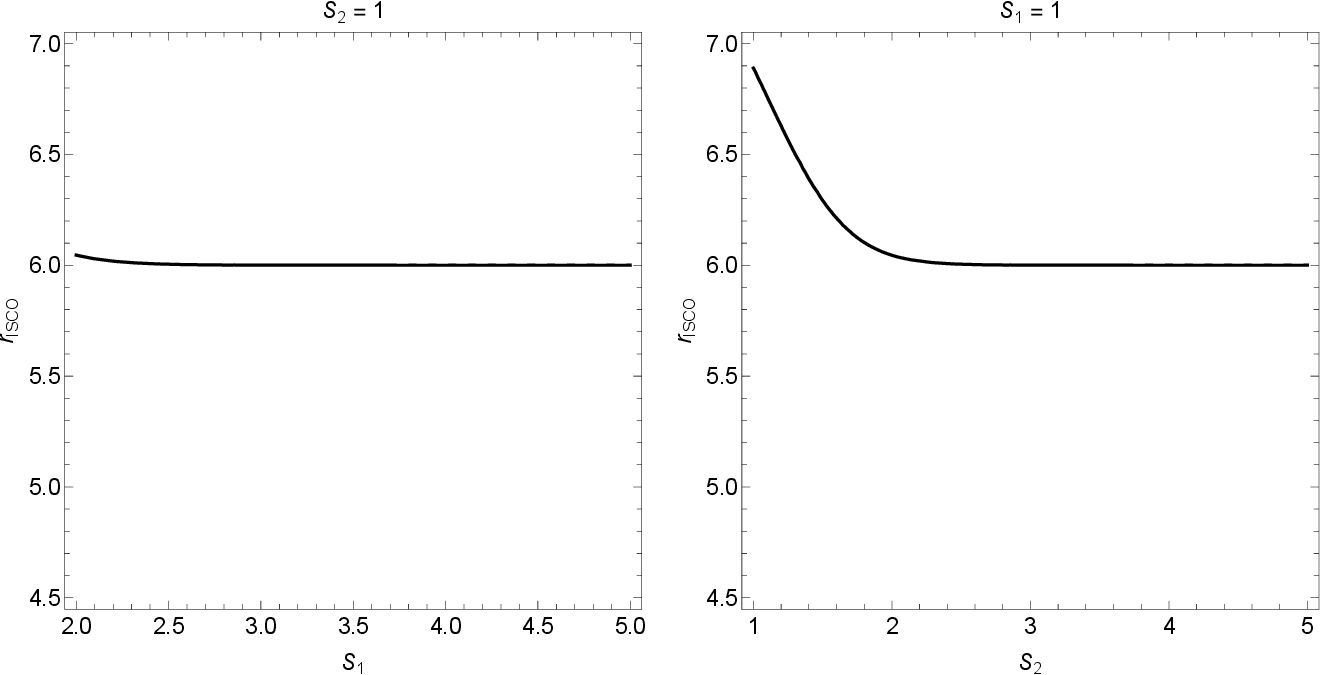}
\caption{Plots for ISCOs of particles moving around a non-rotating Lee-Wick BH.
}\label{fig_ISCO}
\end{figure*}

The locations of stable and unstable circular orbits correspond to the minimum and maximum of the effective potential, respectively. In Newtonian theory, the effective potential always has a minimum for any given angular momentum, and there are no ISCOs with a minimum radius. However, when the form of the effective potential depends on the angular momentum of the particle and other parameters, the situation changes. In general relativity, for particles orbiting near a Schwarzschild BH, the effective potential exhibits two extrema for any angular momentum value. Only for a specific value of the angular momentum do these two points coincide, marking the ISCOs, which is located at $r = 3r_g$, where $r_g$ is the Schwarzschild radius. The ISCOs can be determined by applying certain conditions
\beq
V_{\rm eff}(r) = \EE^2, \quad \frac{\d V_{\rm eff} (r)} {\d \, r} = 0, \quad \frac{\d^2 V_{\rm eff} (r)} {\d\, r^2} = 0.\label{iscos}
\eeq
One can get the radii of ISCOs from the last relation around a non-rotating Lee-Wick BH. Fig. (\ref{fig_ISCO}) illustrates the equatorial ISCOs around a non-rotating Lee-Wick BH. The first panel shows how the ISCOs change for different values of the parameter $S_1$, while the second panel depicts the influence of the parameter $S_2$. As the value of parameter $S_1$ increases, initially, the radii of ISCOs decrease and then become constant. Similarly, increasing the parameter $S_2$ leads to a decrease in the radii of the ISCOs initially and then becomes constant.


\section{Harmonic oscillations as perturbation of circular orbits} \label{oscillations}

To study the oscillatory motion of neutral particles, we perturb the equations of motion around stable circular orbits. When a test particle is slightly displaced from its equilibrium position associated with a stable circular orbit in the equatorial plane, it exhibits epicyclic motion, defined by linear harmonic oscillations. 

\subsection{Frequencies measured by local observer} 

The frequencies of harmonic oscillatory motion determined by the local observer are given by
\bea\label{Freq-2}
\omega_{r}^{2} &=&  \frac{-1}{2} \frac{\partial^{2} V_{\rm eff} (r, \theta)}{\partial r^{2}},\\\label{Freq-3}
\omega_{\theta}^{2} &=& \frac{1}{2}\frac{g_{rr} (r, \theta)}{r^2} \frac{\partial^{2} V_{\rm eff}(r, \theta)}{\partial \theta^{2}},\\
\omega_\phi &=& \frac{\d \phi}{\d \tau}.
\eea
The radial $(\omega_{r})$, latitudinal ($\omega_{\theta}$), and orbital/axial ($\omega_{\phi}$) frequencies of the neutral test particle for a non-rotating Lee-Wick BH takes the form
\begin{widetext}
\bea\non
\omega_{r}^{2} &=& \frac{e^{-r S_1}}{2 r^5 S_1 S_2} [2 S_1 S_2 \left(2 r^2-3 L^2 (r-4)\right) e^{r S_1}-(\LL^2 (r^3 S_1^5+5 r^2 S_1^4+2 r S_1^3 \left(r^2 S_2^2+6\right)+2 S_1^2 \left(5 r^2 S_2^2+6\right) \\\non &+& r S_2^2 S_1 \left(r^2 S_2^2+12\right)+S_2^2 \left(5 r^2 S_2^2-12\right))+r^2 (r^3 S_1^5+2 S_1^3 \left(r^3 S_2^2+r\right) +r^2 S_1^4+2 S_1^2 \left(r^2 S_2^2+1\right) \\\non &+& r S_2^2 S_1 \left(r^2 S_2^2+2\right)+S_2^2 \left(r^2 S_2^2-2\right))) \sin \left(r S_2\right)  + S_2 (\LL^2 \left(r^3 S_1^4 + 2 r S_1^2 \left(r^2 S_2^2-6\right) + r S_2^2 \left(r^2 S_2^2-12\right)-24 S_1\right) \\\label{omega_r} &+& r^2 \left(r^3 S_1^4+2 r S_1^2 \left(r^2 S_2^2-1\right)+r S_2^2 \left(r^2 S_2^2-2\right)-4 S_1\right)) \cos \left(r S_2\right)],\\\non
\omega_{\theta}^{2} &=& \frac{-1}{r^2 (\chi_1 + \chi_2)} [\left(r^2 S_2^4+S_2^2 \left(r S_1+1\right) \left(2 r S_1-1\right)+r S_1^3 \left(r S_1+1\right)+S_1^2\right) \sin \left(r S_2\right)-2 S_1 S_2 e^{r S_1} \\\label{omega_theta} &+& S_2 \left(r S_2^2+S_1 \left(r S_1+2\right)\right) \cos \left(r S_2\right)],\\\non
\omega_\phi^2 &=& \frac{-1}{r^2 (\chi_1 + \chi_2)} [\left(r^2 S_2^4+S_2^2 \left(r S_1+1\right) \left(2 r S_1-1\right)+r S_1^3 \left(r S_1+1\right)+S_1^2\right) \sin \left(r S_2\right)-2 S_1 S_2 e^{r S_1} \\\label{omega_phi} &+& S_2 \left(r S_2^2+S_1 \left(r S_1+2\right)\right) \cos \left(r S_2\right)],
\eea
\end{widetext}
where
\begin{eqnarray*}
\chi_1 &&=2 (r-3) S_1 S_2 e^{r S_1}+3 S_2 \left(r S_2^2+S_1 \left(r S_1+2\right)\right)\\&&\times \cos \left(r S_2\right).\\
\chi_2 &&=\left(r^2 S_2^4+S_2^2 \left(r S_1 \left(2 r S_1+3\right)-3\right)+S_1^2 \right. \\ &&\times \left. \left(r S_1 \left(r S_1+3\right)+3\right)\right) \sin \left(r S_2\right).
\end{eqnarray*}

\begin{figure*}
\centering 
\includegraphics[width=\hsize]{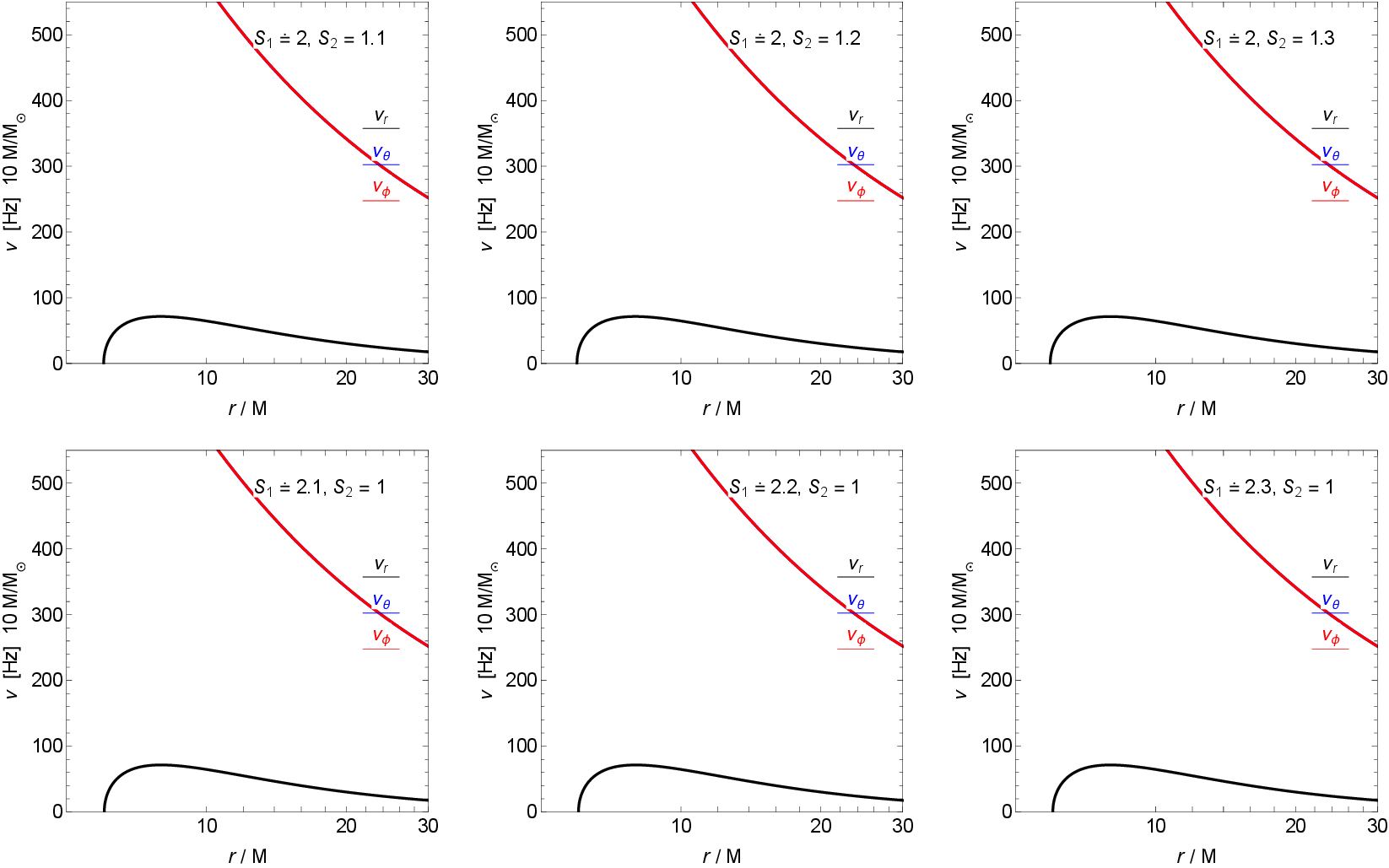}
\caption{Plots of frequencies of particles moving around a non-rotating Lee-Wick BH.
}\label{fig_frequencies}
\end{figure*}

\begin{figure*}
\centering 
\includegraphics[width=\hsize]{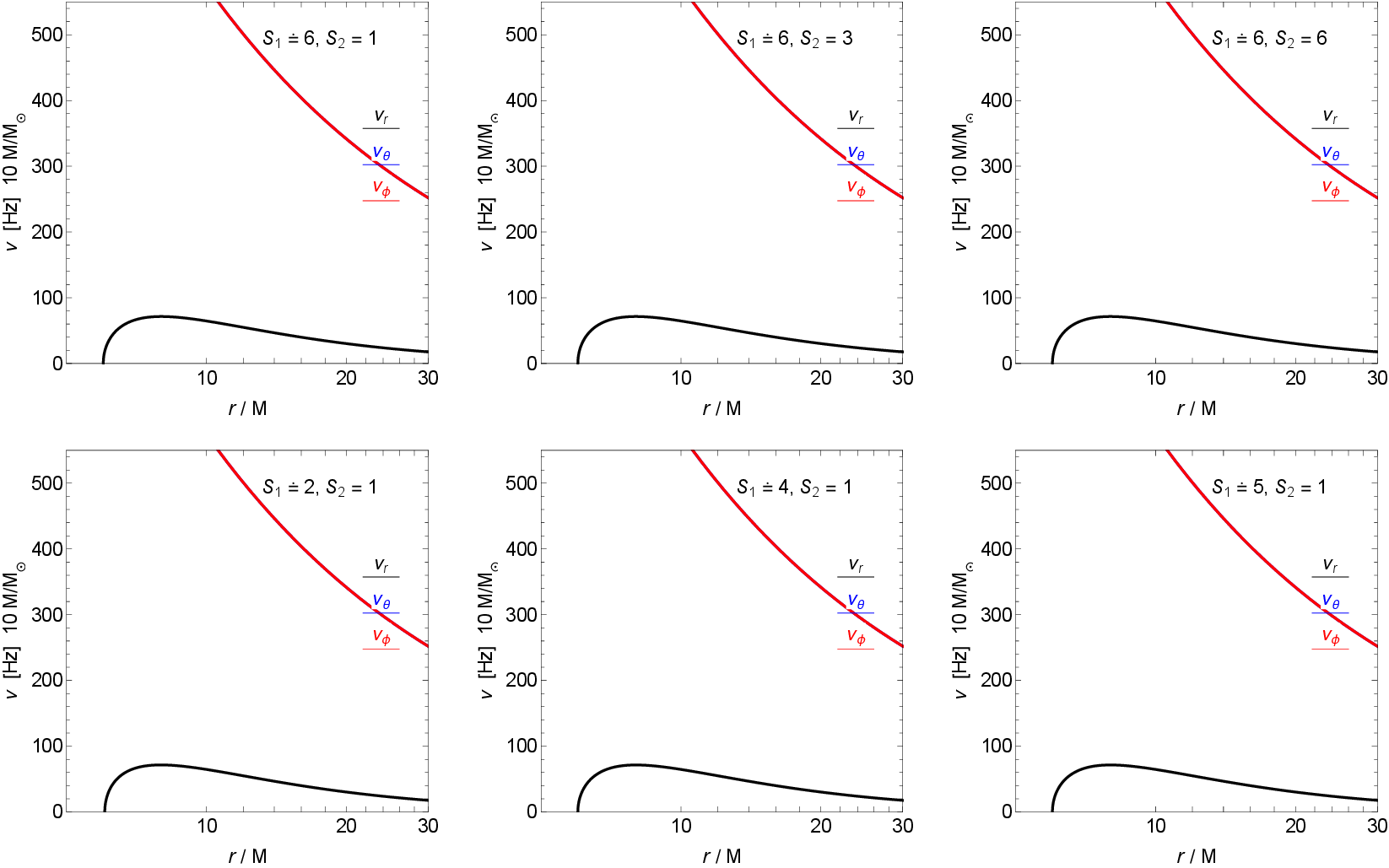}
\caption{Plots of frequencies of particles moving around a non-rotating Lee-Wick BH.
}\label{fig_frequencies1}
\end{figure*}

\subsection{Frequencies measured by distant observer} 

The locally determined angular frequencies $\omega_\beta$ are specified in Eqs.~((\ref{omega_r})-(\ref{omega_phi})). In contrast, the angular frequencies measured by a static distant observer ($\Omega$) are given by
\beq\label{frequencies}
\Omega_{\beta} = \omega_{\beta} \frac{\d \tau}{\d t},
\eeq
where $\tau/\d t$ is the redshift coefficient, given by
\begin{equation}
    \frac{\d t}{\d \tau} = - \frac{\EE }{g_{tt}}.
\end{equation}
If the frequencies of small harmonic oscillations are measured in physical units by a distant observer, then the frequencies of neutral particles determined by these distant observers take the form
\beq\label{nu_rel}
\nu_{i}=\frac{1}{2\pi}\frac{c^{3}}{GM} \, \Omega_{i}[{\rm Hz}],
\eeq
where $i\in\{r,\theta,\phi\}$; $\Omega_{r}$, $\Omega_{\theta}$, and $\Omega_{\phi}$ denote the dimensionless radial, latitudinal, and axial angular frequencies measured by a distant observer, given in the Appendix. 

Figs. (\ref{fig_frequencies}) and (\ref{fig_frequencies1}) present the radial profiles of the frequencies $\nu_{j}$ for small harmonic oscillations of neutral particles around a non-rotating Lee-Wick BH, observed by a distant observer. Different cases of parameters are plotted. The radial profiles of the orbital frequency ($\Omega_{\phi}$) and the latitudinal frequency ($\Omega_{\theta}$) coincide. As the BH parameter $S_2$ increases, the profiles shift closer to the event horizon of the BH. In particular, an increase in the parameter $S_1$ also causes the profiles to move towards the BH horizon.

\begin{figure*}
\centering 
\includegraphics[width=\hsize]{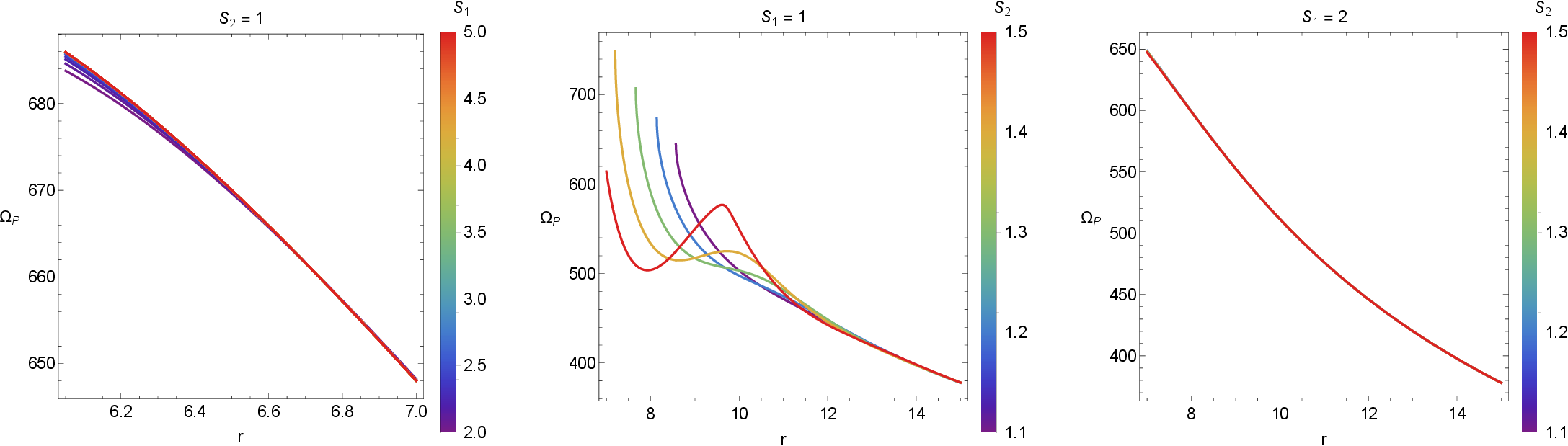}
\caption{Plots for periastron frequency of particles around non-rotating Lee-Wick BH.
}\label{precession}
\end{figure*}

\subsection{Periastron precession} 

In this section, we examine the periastron frequency of a neutral test particle orbiting a non-rotating Lee-Wick BH, focusing on small perturbations around the equatorial plane at $\pi/2$. To calculate the periastron precession, we consider a slight perturbation of the particle from its stable position, which leads to oscillations around that stable point characterized by a radial frequency $\Omega_{r}$. The periastron frequency, denoted $\Omega_{P}$, is defined as the difference between the orbital frequency $\Omega_{\phi}$ and the radial frequency $\Omega_{r}$, as expressed in the following relation
\beq
\Omega_{P} = \Omega_{\phi} - \Omega_{r},
\eeq
Unlike Newtonian gravity, where the radial and orbital frequencies are equal, general relativistic effects near a BH lead to the inequality $\Omega_{\theta} \neq \Omega_{\phi}$. Fig. (\ref{precession}) shows the graphical behavior of the periastron frequency for particles orbiting a non-rotating Lee-Wick BH, presenting the effects of different BH parameters $S_1$ and $S_2$. The first column is plotted for different values of the parameter $S_1$. As the parameter $S_1$ increases, the frequency of the periastron increases. The second and third columns are plotted for different values of the BH parameter $S_2$. For $S_1=1$, different radial profiles can be observed for different values of the BH parameter $S_2$. However, for $S_1=2$, different profiles coincide.  

\section{Numerical Simulations}\label{NumSim} 

In this section, in addition to the theoretical developments, we investigate the dynamic structure of the shock cone produced by BHL accretion around a Lee-Wick BH, as well as the modifications induced by Lee-Wick gravity, by numerically solving the general relativistic hydrodynamics equations. To this end, we adapt the numerical code used in our previous work \cite{Orh2,Orh3,Orh4,Orh5,Orh6,Orh7,Orh8}, configuring it with the appropriate initial and boundary conditions. The plasma distribution and the shock cone that form around the Lee-Wick BH are initialized using the parameter sets provided in Tab. (\ref{Inital_Con_1}). Each row of Tab. (\ref{Inital_Con_1}) lists the model name, the Lee-Wick parameters \(S_1\) and \(S_2\), the inner and outer horizons of the BH for that model, and the inner radial boundary of the computational domain (that is, the radius of the innermost disk closest to the BH horizon). In the first block of Tab. (\ref{Inital_Con_1}), we model cases where the event horizon is close to \(2M\) or slightly smaller.  Consequently, the flow yields Schwarzschild-like outcomes, and below we present numerical results together with comparisons to the Schwarzschild case. In contrast, the second block considers cases with an event-horizon radius \(>2M\) as illustrated in Fig. (\ref{fig_LapseFun}), especially for \(S_1=1\), where the altered behavior of the lapse function manifests itself in plasma morphology and in the formation of the cone and bow shock.

In the non-rotating Lee-Wick BH metric given by Eqs.\ref{metric} and \ref{metric2}, the lapse function \(f(r)\) depends on the parameters \(S_1\) and \(S_2\). The parameter \(S_1\) sets the exponential decay scale of the Lee-Wick correction, while \(S_2\) controls its oscillatory behavior. As shown in Fig. (\ref{fig_LapseFun}), the zeros of the lapse, \(f(r)=0\), determine the inner and outer horizons, which are listed in Tab. (\ref{Inital_Con_1}) for the corresponding pairs \((S_1,S_2)\). In the metric, one finds that for \(r \gg 1/S_1\) the spacetime approaches the Schwarzschild limit, and the sensitivity of \(f(r)\) to \(S_2\) diminishes as \(S_1\) increases (as seen in the lapse-function sweeps). In contrast, for \(S_1=1\) there is a pronounced variation with \(S_2\), whereas for \(S_1=2\) almost no change is observed, since the event horizons form at nearly the same radii as in the Schwarzschild case. Therefore, the choice of \((S_1,S_2)\) significantly modifies the shock cone morphology produced by BHL accretion. Oriented by the lapse behavior in Fig. (\ref{fig_LapseFun}) and the two initial data blocks listed in Tab. (\ref{Inital_Con_1}), we performed the numerical simulations accordingly.

\setlength{\tabcolsep}{12pt}
\begin{table*}[htbp]
\footnotesize
\caption{The table shows the horizon structure of the non-rotating Lee-Wick BH and the values of the parameters $S_1$ and $S_2$ used in the numerical simulation. In the table, $r_+$ represents the outer horizon, while $r_-$ denotes the inner horizon. ``NO'' indicates that no horizon exists for the corresponding parameter values.
}
 \label{Inital_Con_1}
\begin{center}
  \begin{tabular}{c|c|c|c|c|c}
    \hline
    \hline
    Model & $S_1 (\frac{1}{M})$ & $S_2 (\frac{1}{M})$ & $r_+(M)$ &  $r_-(M)$ & $r_{min}(M)$ \\    
   \hline  
                 & $1$ & $1$ & $NO$ & $NO$ & $2.3$ \\
          & $6$ & $1$ & $1.99958$ & $0.1647$ & $2.3$\\   
                  & $6$ & $3$ & $1.999454$ & $0.125$ & $2.3$ \\
        $Block-1$          & $6$ & $6$ & $1.999999$ & $0.06$ & $2.3$\\   
         & $2$ & $1$ & $1.6858$ & $1.1944$ & $2.3$\\    
          & $4$ & $1$ & $1.989477$ & $0.33011$ & $2.3$\\      
          & $5$ & $1$ & $1.999876$ & $NO$ & $2.3$\\    
   \hline
          & $1$ & $1.5$ & $2.719$ & $0.888$ & $3$\\     
   $Block-2$        & $1$ & $2$ & $2.372058$ & $0.418659$ & $3$ \\ 
          & $1.5$ & $1.5$ & $2.330$ & $0.754$ & $3$ \\    
          & $1.5$ & $4$ & $2.49963$ & $0.21396/1.31999/2.01775$ & $3$ \\      
    \hline
    \hline
  \end{tabular}
\end{center}
\end{table*}
%

\subsection{\text{Block-1}: Weak Lee-Wick Regime}
\label{B1} 
 
In the \text{Block-1} models presented in Tab. (\ref{Inital_Con_1}), only the values of $S_{1}$ and $S_{2}$ that produce weak correction terms around the Lee-Wick BH are considered. These parameters lead to only small deviations from the classical Schwarzschild geometry. In this domain, the exponential suppression controlled by $S_{1}$ dominates over the oscillatory contribution generated by $S_{2}$, so the metric describing Lee-Wick gravity preserves the essential features of general relativity and remains nearly unchanged. Only higher-derivative terms introduce mild modifications in the vicinity of the BH horizon. The spacetime metric remains smooth and regular, and the position of the horizon exhibits only minimal shifts compared to the Schwarzschild case. Therefore, this region corresponds to the near-classical limit of the Lee-Wick model, providing a consistent setting for examining how subtle quantum-correction terms modify the spacetime geometry. In this section, we reveal how these small corrections influence the morphology of the shock cone and the characteristics of the QPOs obtained from the numerical simulations.

\subsubsection{Morphology of Accretion Flow and Shock-Cone Dynamics}\label{Morph_B1} 
\begin{figure*}
\centering 
\includegraphics[width=7.5cm,height=5.9cm]{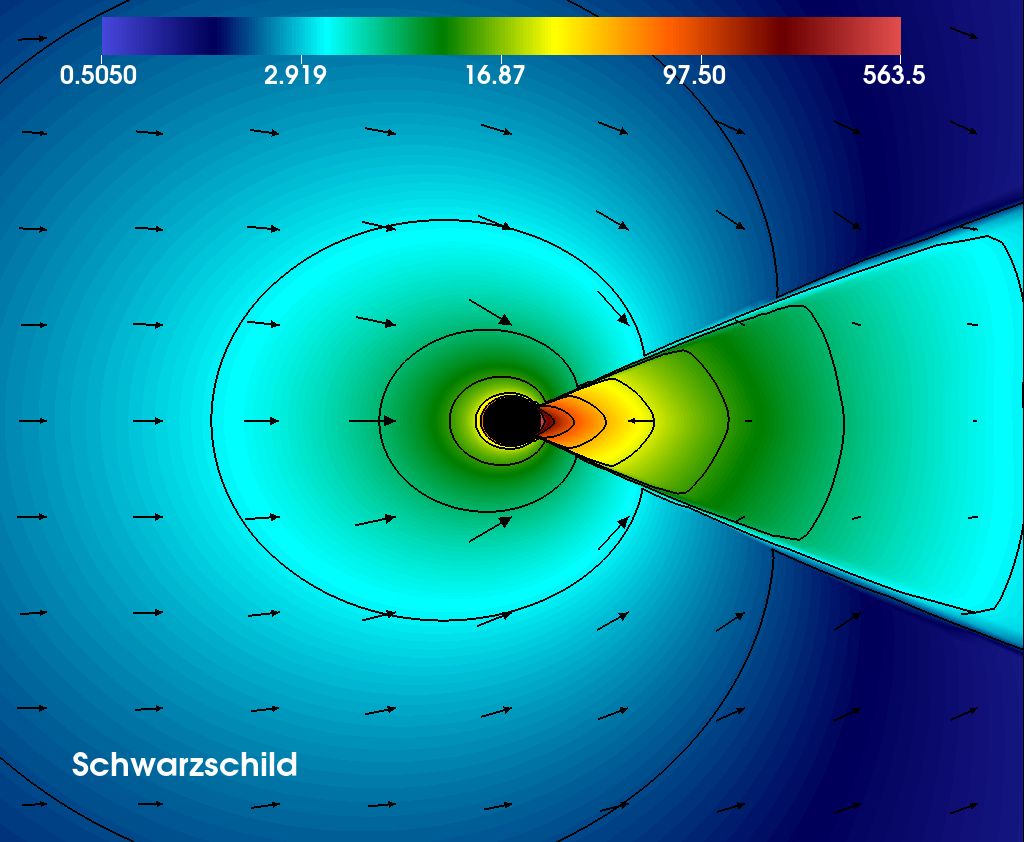} 
\includegraphics[width=7.5cm,height=5.9cm]{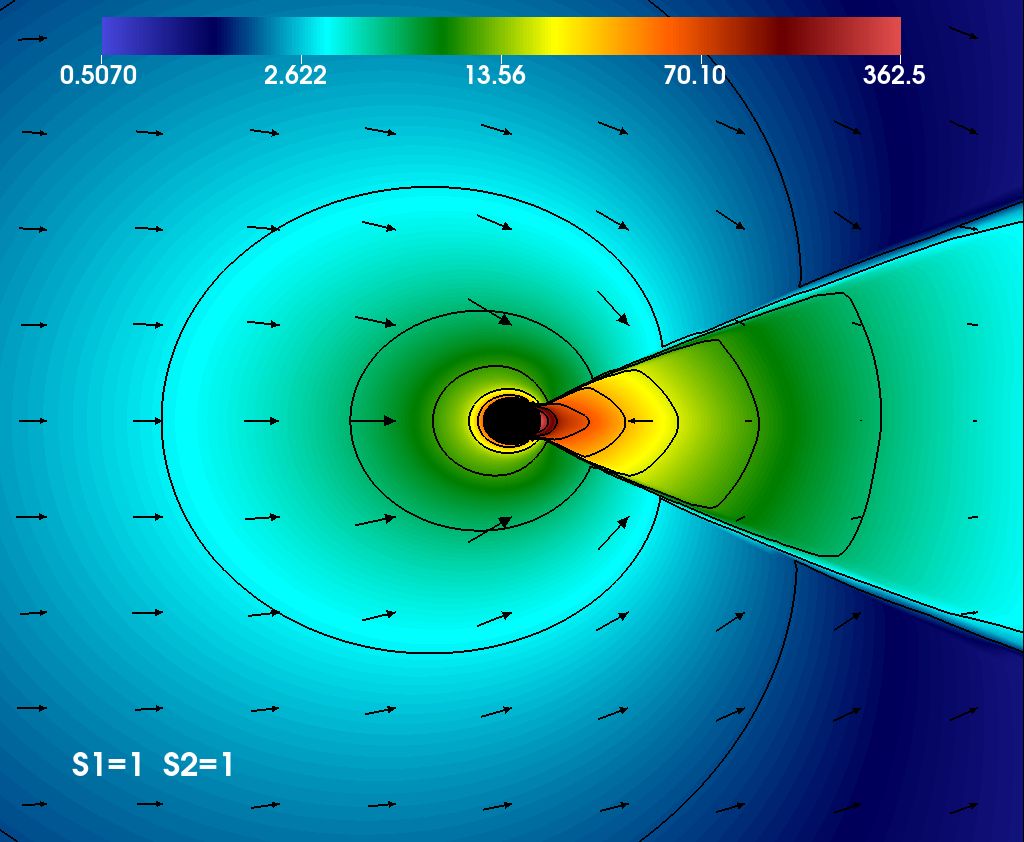}\\
\includegraphics[width=7.5cm,height=5.9cm]{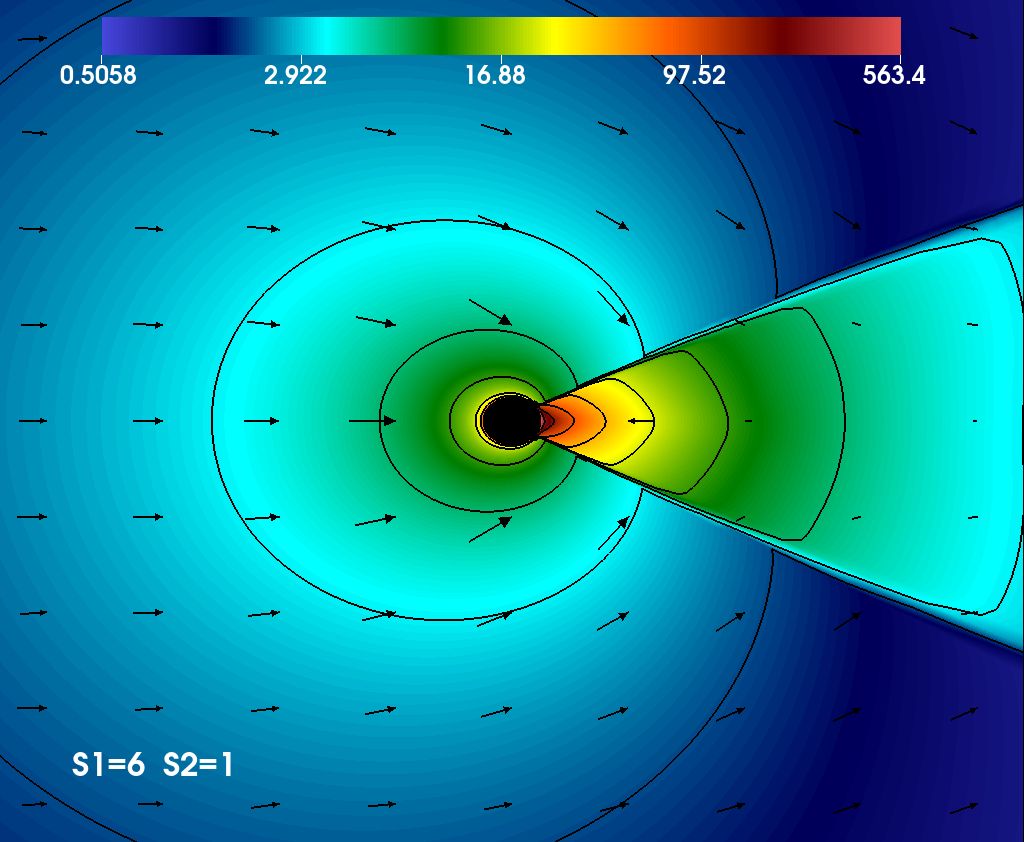} 
\includegraphics[width=7.5cm,height=5.9cm]{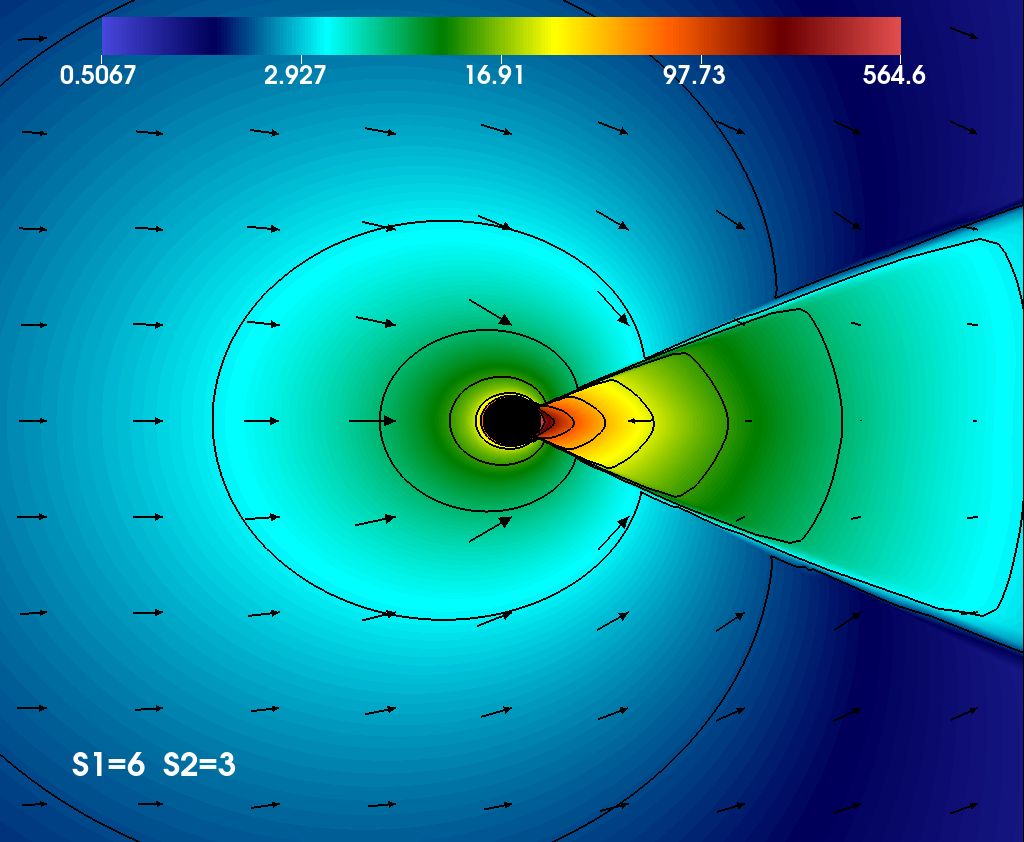}\\
\includegraphics[width=7.5cm,height=5.9cm]{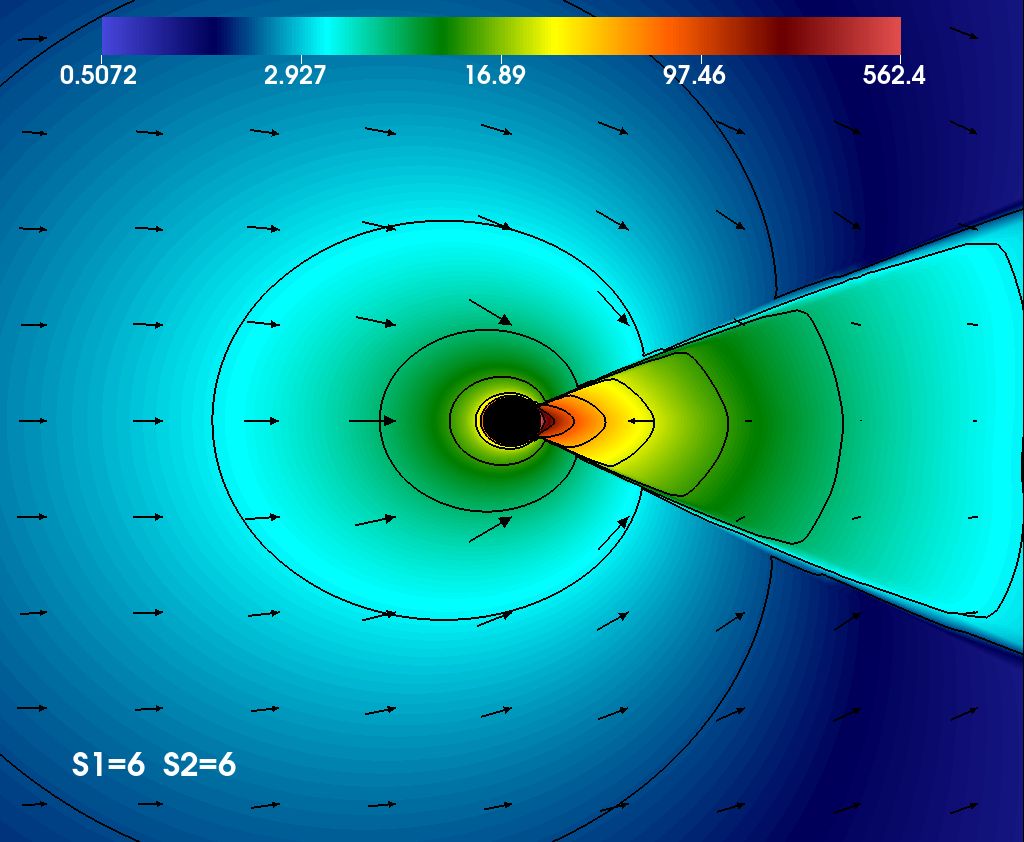} 
\includegraphics[width=7.5cm,height=5.9cm]{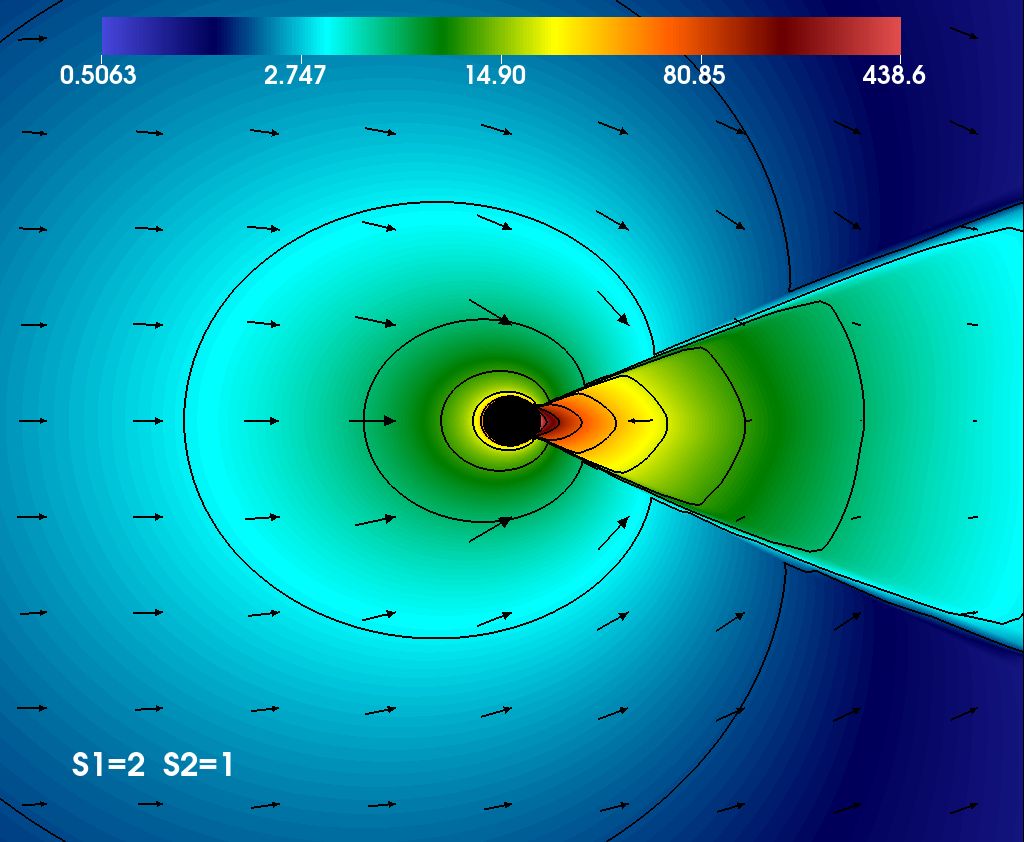}\\
\includegraphics[width=7.5cm,height=5.9cm]{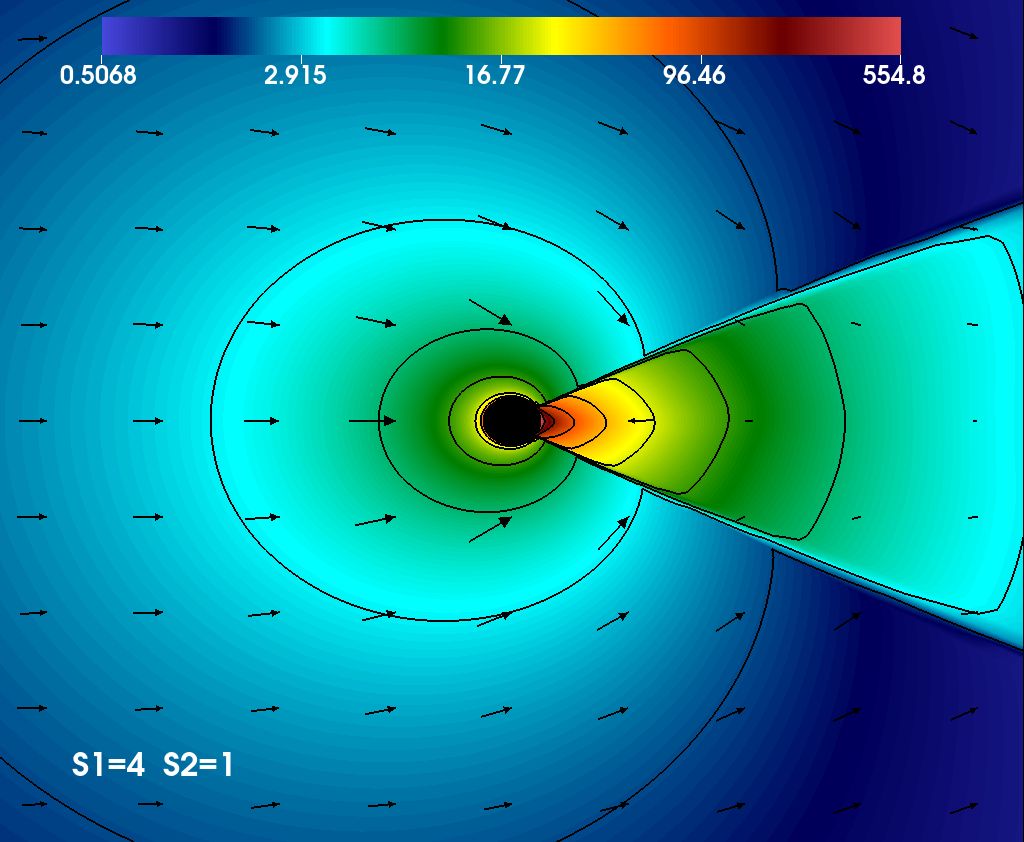} 
\includegraphics[width=7.5cm,height=5.9cm]{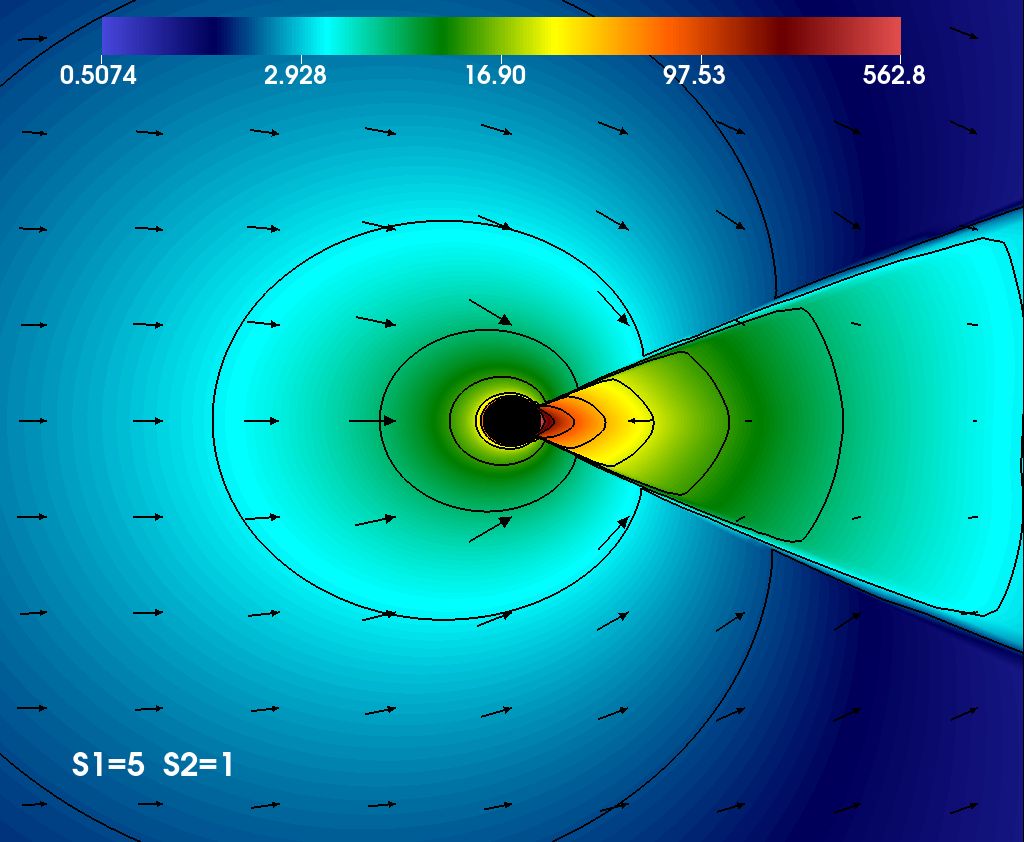}
\caption{In the \text{Block-1} case given in  Table \ref{Inital_Con_1}, together with the Schwarzschild solution, we plot the rest-mass density in the region close to the BH along with its contour map and vector (velocity) field on the equatorial plane to display the morphology of the resulting shock cone. 
}\label{color_block1}
\end{figure*}

\begin{figure*}
\centering 
\includegraphics[width=15.0cm,height=12.0cm]{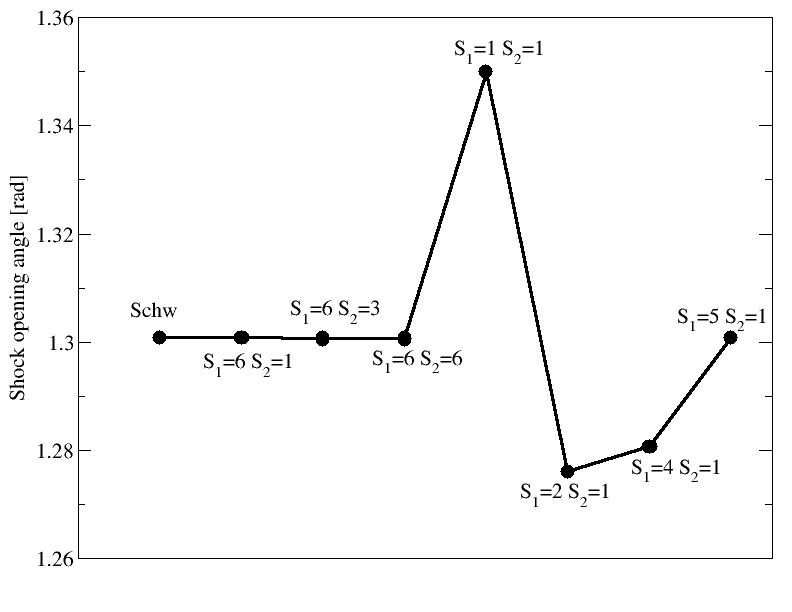} 
\caption{In the \text{Block-1} configuration, we present the dependence of the shock cone opening angle on \(S_1\) and \(S_2\), shown alongside the Schwarzschild case. Consistent with theoretical expectations, at sufficiently large \(S_1\) the opening angle converges to the Schwarzschild result.  
}\label{shock_ang_Block1}
\end{figure*}

Understanding how the morphology of the shock cone formed in the downstream region of the BHL accretion flow changes with the Lee-Wick parameters $S_{1}$ and $S_{2}$ around a nonrotating BH helps reveal their influence on both the accretion structure and the resulting QPOs. For this purpose, Fig.\ref{color_block1} presents the dynamic structure of the shock cone obtained for the \text{Block-1} parameter sets, together with the corresponding case of a Schwarzschild BH for direct comparison. The figure shows the rest-mass density distribution of the accreting matter on the equatorial plane. Both contour and vector plots are included for each model, allowing the behavior of the flow to be examined once the cone reaches a steady state. In all cases, the system clearly attains a quasi-steady configuration. The weak correction regime, where $S_{1}$ and $S_{2}$ introduce only mild deviations from the Schwarzschild spacetime, yields an overall shock cone morphology that is almost identical to the classical BHL configuration. The cone formed in the downstream region is smooth, symmetric, and tightly bound to the inner boundary of the computational domain at $r = 2.3M$. This demonstrates that the plasma dynamics are primarily governed by the standard general relativistic potential rather than by the higher derivative corrections of Lee-Wick gravity. However, as $S_{2}$ increases, the density contours along the cone begin to show slight asymmetry, and the opening angle of the cone becomes marginally smaller. The dependence of this angle on the parameters, shown in Fig. (\ref{shock_ang_Block1}) and discussed later in detail, is consistent with the exponential oscillatory competition appearing in the lapse function $f(r)$ given by Eq.\ref{metric2}. The numerically obtained shock morphologies in the weak Lee Wick regime agree fully with the theoretical expectations derived from the effective potential profiles presented in Fig. (\ref{fig_Veff}).  In particular, larger values of $S_{1}$ produce strong exponential suppression in $f(r)$, deepening the gravitational potential well and creating a more compact, high-density region near $r \simeq 2M$. Meanwhile, increasing $S_{2}$ induces oscillatory modulation in both $f(r)$ and the curvature terms, slightly perturbing the equilibrium structure without destabilizing the flow trapped within the cone. 

\begin{figure*}
\centering 
\includegraphics[width=15.0cm,height=11.5cm]{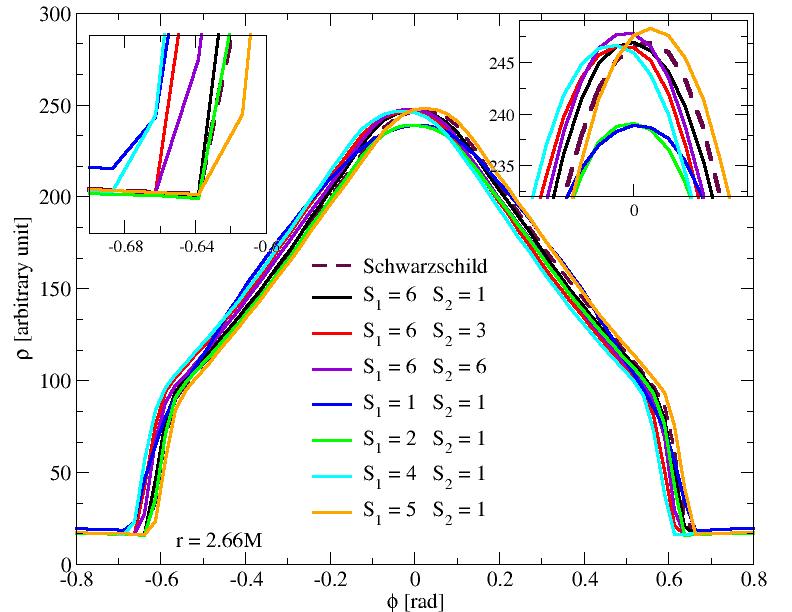} \\
\includegraphics[width=15.0cm,height=11.5cm]{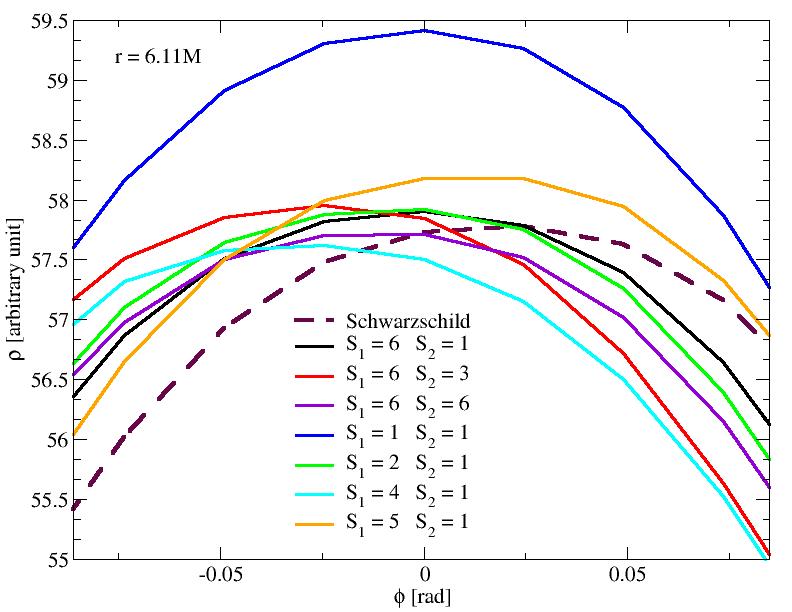} 
\caption{It shows the dynamic structure of the shock cone formed around the non-rotating Lee-Wick BH and the parameter-dependent variation of the rest-mass density trapped inside the cone along the azimuthal direction at two different radial positions. In the upper panel, the variation at $r = 2.66M$, the closest point to the BH, is presented for different values of $S_1$ and $S_2$ given in the \text{Block-1} in comparison with the Schwarzschild case. The small insets within the figure display the dependence of the strong shock region and the maximum density on the parameters.  In contrast, the lower panel shows how the maximum density of matter trapped inside the cone varies with the parameters at $r = 6.11M$, i.e., near the ISCOs, also compared with the Schwarzschild case.
}\label{azimutal_den}
\end{figure*}

In the weak-correction regime, the azimuthal variation of the rest-mass density inside the shock cone formed by the BHL accretion mechanism around the nonrotating Lee-Wick BH is illustrated in Fig. (\ref{azimutal_den}). The top panel shows this variation at $r = 2.66M$, very close to the BH horizon, while the bottom panel presents the corresponding distribution near the ISCOs region. The rest-mass density profiles at both radial points are compared with the Schwarzschild case to demonstrate the dependence of the shock cone morphology on the Lee-Wick parameters $S_1$ and $S_2$ within the weak field regime. At $r = 2.66M$, the cone is compact and well developed. The slight deviations observed for different combinations of $(S_1, S_2)$ arise from the modifications of gravity induced by the Lee-Wick corrections.  As the parameter $S_1$, which controls the exponential suppression term in the lapse function $f(r)$, increases, the resulting deeper effective potential well leads to a more dense and confined plasma region near the cone axis. In contrast, an increase in $S_2$, which regulates the oscillatory component of $f(r)$, generates small azimuthal fluctuations and introduces a slight asymmetry compared to the Schwarzschild case. The insets inside the top panel clearly display the dependence of both the maximum density and the width of the strong shock region on the Lee-Wick parameters. For large values of $S_1$ ($S_1 \geq 2$), these variations converge towards the Schwarzschild solution, indicating that strong exponential damping suppresses the higher derivative corrections.  

\begin{figure*}
\centering 
\includegraphics[width=15.0cm,height=12.0cm]{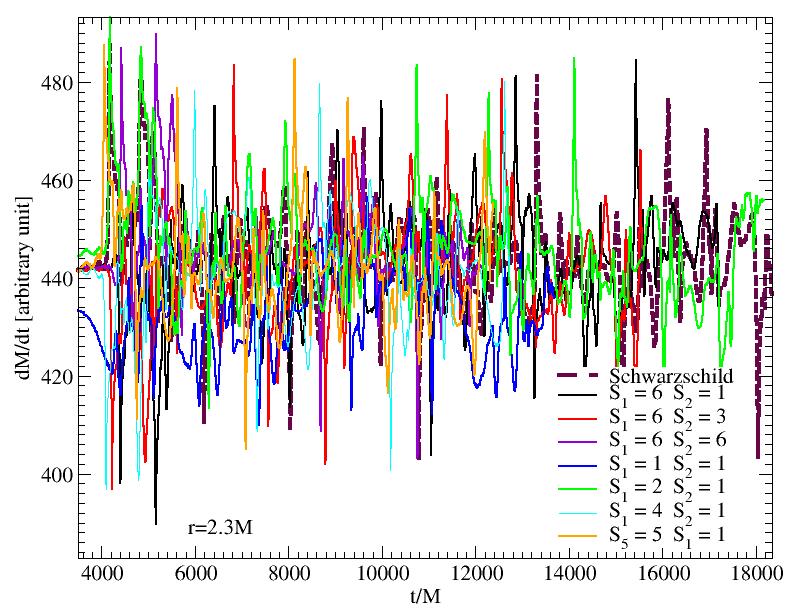} 
\caption{Time evolution of the mass accretion rate at $r = 2.3M$ around the non-rotating Lee-Wick BH for different values of $S_1$ and $S_2$ given in the \text{Block-1}, shown together with the Schwarzschild case for comparison.  The figure illustrates the strong instabilities in the plasma structure, which create a shock cone surrounding the BH and the overall inflow rate of matter onto it.
}\label{massAcc}
\end{figure*}

In the bottom panel of Fig. (\ref{azimutal_den}), the azimuthal variation of the density of the mass of the rest is shown in $r = 6.11M$, corresponding to the region where the flow transitions from purely radial accretion to rotational motion near the ISCOs. Compared with the $r = 2.66M$ case, the density amplitude decreases in all models due to the weaker gravitational confinement at larger radii. However, the dependence on $S_1$ and $S_2$ follows the same trend observed in the top panel. The larger $S_1$ values produce a tighter and more confined shock cone, while the higher $S_2$ values induce fine-scale oscillations in the azimuthal density profile. This region is particularly important since QPOs can emerge here, linking the spatial morphology of the cone to the temporal fluctuations in the accretion rate, as shown later in Fig. (\ref{massAcc}). Compared to the Schwarzschild case, both radial positions reveal that the Lee-Wick parameters $S_1$ and $S_2$ produce observable modifications in the shock cone structure while maintaining a symmetric and smooth overall geometry. The numerical results obtained here are in excellent agreement with the theoretical predictions derived earlier in the paper.  

In addition to Fig. (\ref{azimutal_den}), Fig. (\ref{shock_ang_Block1}) illustrates how the opening angle of the shock cone varies with the Lee-Wick parameters. As seen in Fig. (\ref{shock_ang_Block1}), increasing $S_1$ causes the cone opening angle to gradually converge to the value obtained for the Schwarzschild solution. This confirms that the strong exponential suppression term minimizes deviations from general relativity, consistent with the theoretical discussion above. Meanwhile, variations in $S_2$ produce noticeable changes in cone width, reflecting the interaction between oscillatory and exponential terms in the lapse function. For moderate values of $S_1$ and $S_2$, the smaller opening angle indicates a more compact inflow structure produced by stronger gravitational focusing. This result is consistent with the improved compression and narrow cone geometry shown in Fig. (\ref{azimutal_den}). Altogether, these findings confirm that even mild quantum gravity corrections can leave subtle yet observable imprints on the accretion dynamics around the Lee-Wick BH.

When considering the case of $S_1 = 1$ and $S_2 = 1$ in Fig. (\ref{shock_ang_Block1}), we observe analytically that no classical BH horizon is formed, as seen in Tab. (\ref{Inital_Con_1}). In this configuration, the spacetime corresponds to a naked singularity. Consequently, as the accreting matter falls toward the center, the gravitational pull acting on it at $r = 2.3M$ is significantly weaker compared to the regular BH cases. Due to this reduced gravitational acceleration, the inflowing plasma becomes less focused and tends to spread more before reaching the downstream region. As a result, a shock cone is formed with a wider opening angle, lower density, and weaker compression. This broader and less confined cone structure is a direct manifestation of the weaker gravitational confinement around the naked singularity.

In the previous analysis, we demonstrated the formation of a shock cone around the nonrotating Lee-Wick BH and showed that its dynamic structure varies with the model parameters. In this section, we calculate the mass accretion rate in order to investigate the possible emergence of QPOs. For this purpose, the accretion rate is measured at $r = 2.3M$, corresponding to the inner boundary of the computational domain, where the inflowing matter leaves the simulation grid and falls into the BH. By evaluating the time-dependent behavior of $dM/dt$, we determine how the accretion rate changes with respect to the Lee-Wick parameters $S_1$ and $S_2$. Fig. (\ref{massAcc}) shows the temporal evolution of the mass accretion rate at $r = 2.3M$ for the \text{Block-1} models, which include different combinations of $(S_1, S_2)$, together with the Schwarzschild reference case for direct comparison. All models exhibit noticeable temporal variations in the accretion rate, confirming that the accretion flow and the shock cone surrounding the BH are inherently dynamic rather than completely stationary. In the Schwarzschild case, the accretion rate shows small but distinct periodic oscillations, demonstrating that even without any quantum corrections, the downstream shock cone produces moderate and self-sustained instabilities. However, these oscillations remain relatively regular and have lower amplitudes. In contrast, Lee-Wick BH models show stronger and less regular variability in $dM/dt$, with the amplitude and frequency of oscillations showing a mild dependence on the parameters $S_1$ and $S_2$.  

As the parameter $S_1$ increases, the exponential suppression term in the lapse function modifies the effective potential, resulting in a denser inflow of matter and higher amplitude oscillations. On the other hand, an increase in $S_2$ enhances the oscillatory nature of the metric, producing more complex and quasi-periodic modulations that differ from the smoother Schwarzschild pattern. This parameter-dependent behavior indicates that the interplay between the exponential and oscillatory corrections in the Lee-Wick metric destabilizes the post-shock flow, leading to turbulence-like variability and the emergence of QPO-like features. Although the Schwarzschild solution already exhibits intrinsic oscillations associated with the breathing of the shock cone, the Lee-Wick configurations amplify these oscillations and modify their characteristic frequencies. The degree and nature of these changes are discussed in detail in Section \ref{QPOs_B1}. The time-dependent behavior of the accretion rate, particularly for large values of $S_2$, confirms the formation of standing shock wave structures and strong global oscillation modes within the downstream region. These enhanced oscillatory modes could represent potential observational signatures of Lee-Wick gravity. Although these results are obtained under weak-field initial conditions, the differences observed when compared to the Schwarzschild solution suggest that such variability patterns may offer a possible avenue for testing Lee-Wick gravitational effects in future observational or numerical studies.

\subsubsection{Oscillatory Accretion and Quasi-Periodic Signatures }\label{QPOs_B1} 

\begin{figure*}
\centering 
\includegraphics[width=8.0cm,height=7.0cm]{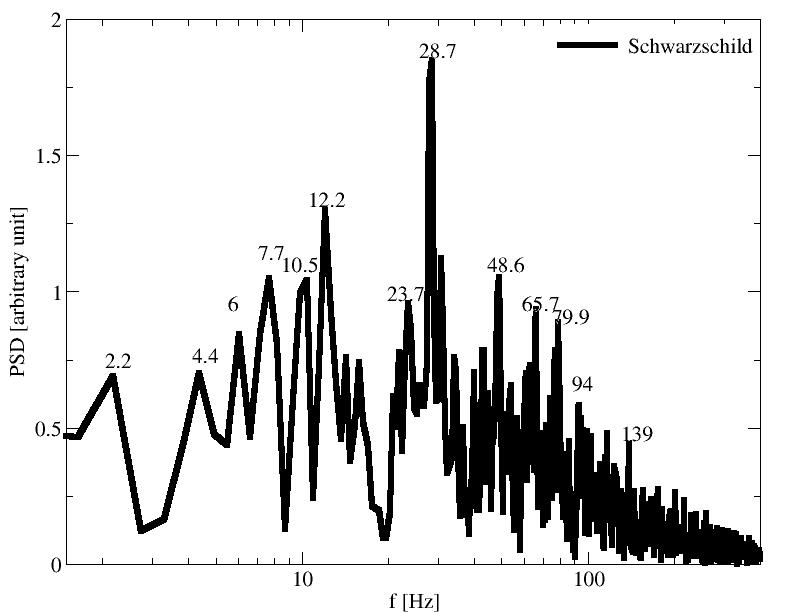} 
\includegraphics[width=8.0cm,height=7.0cm]{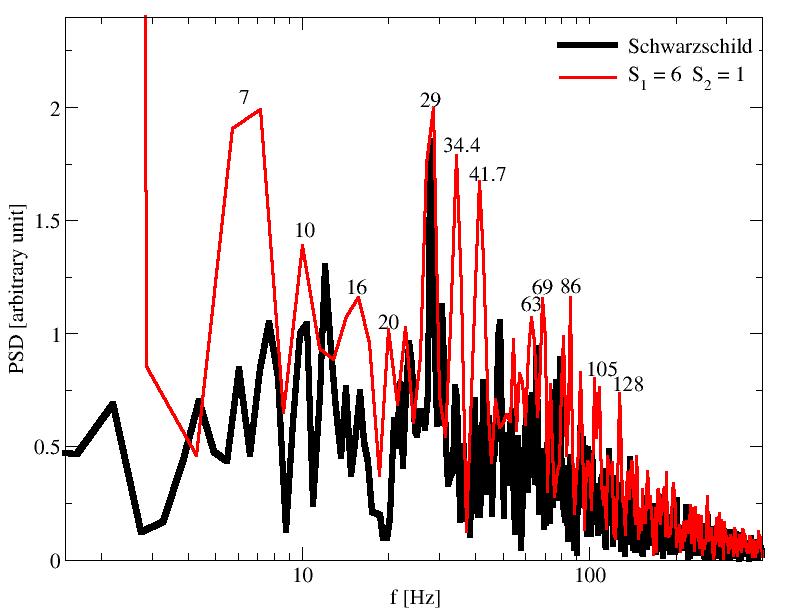}\\ 
\includegraphics[width=8.0cm,height=7.0cm]{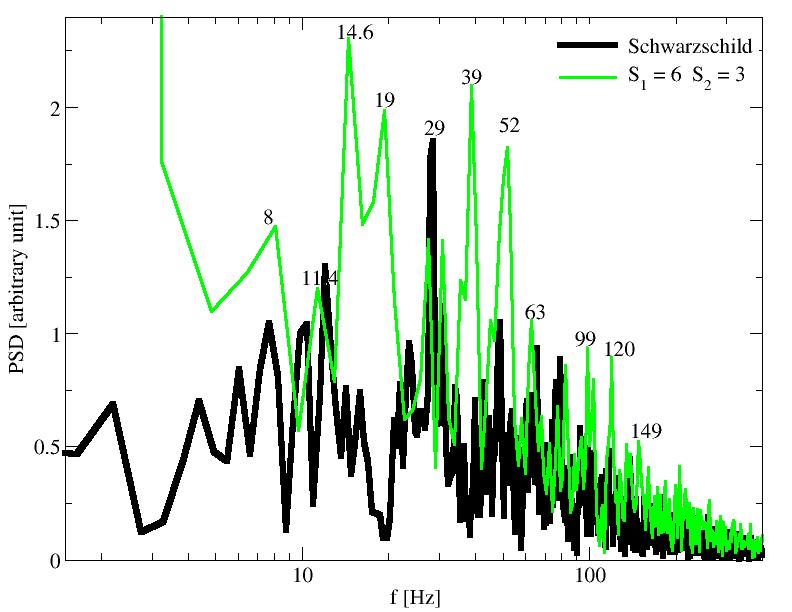}
\includegraphics[width=8.0cm,height=7.0cm]{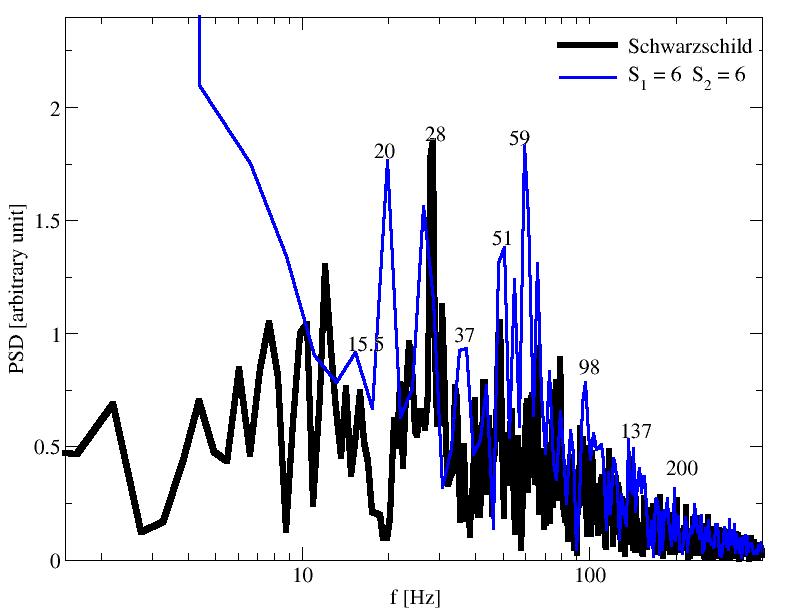}
\caption{The Power Spectral Density analysis is performed at $r = 2.3M$ using the mass accretion rate (Fig.\ref{massAcc}) to reveal the quasi-periodic behavior around both the Schwarzschild and Lee-Wick BHs for different values of $S_1$ and $S_2$ given in the \text{Block-1}.  In the top-left panel, the PSD analysis and the corresponding QPO frequencies are shown only for the Schwarzschild case, while in the other panels both Schwarzschild and Lee-Wick BHs are plotted together. However, the peak frequencies indicated in the PSD correspond only to those that emerge in the Lee-Wick BH cases. As clearly seen from the PSD analyses, increasing values of $S_2$ cause the QPO frequencies to shift from LFQPO to HFQPO regimes.
}\label{QPOs1}
\end{figure*}

\begin{figure*}
\centering 
\includegraphics[width=8.0cm,height=7.0cm]{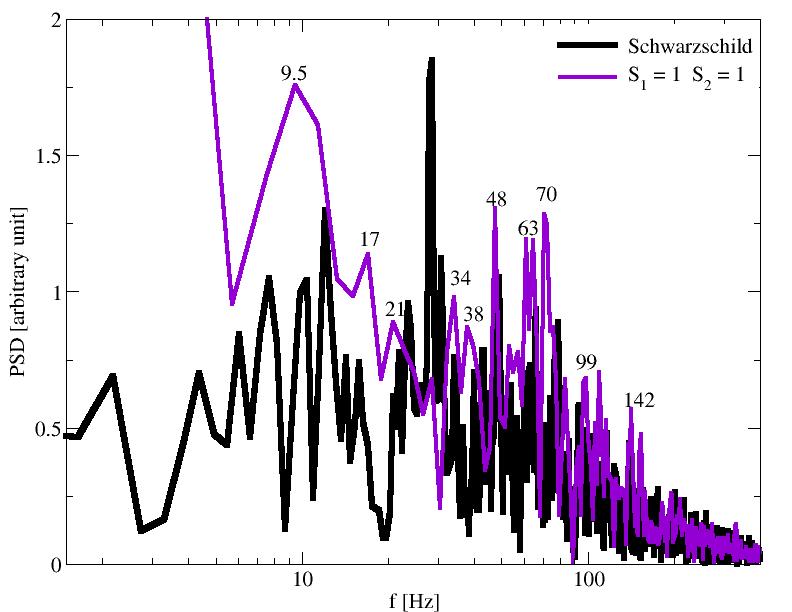} 
\includegraphics[width=8.0cm,height=7.0cm]{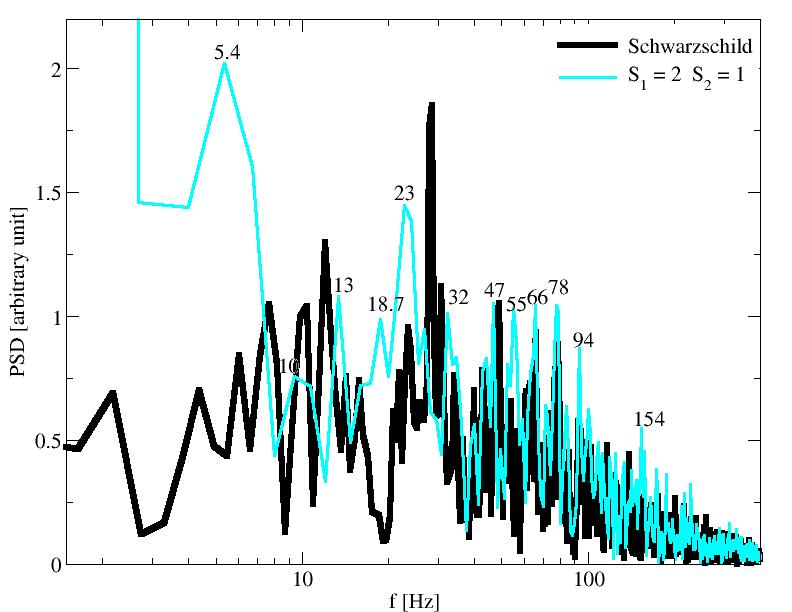}\\ 
\includegraphics[width=8.0cm,height=7.0cm]{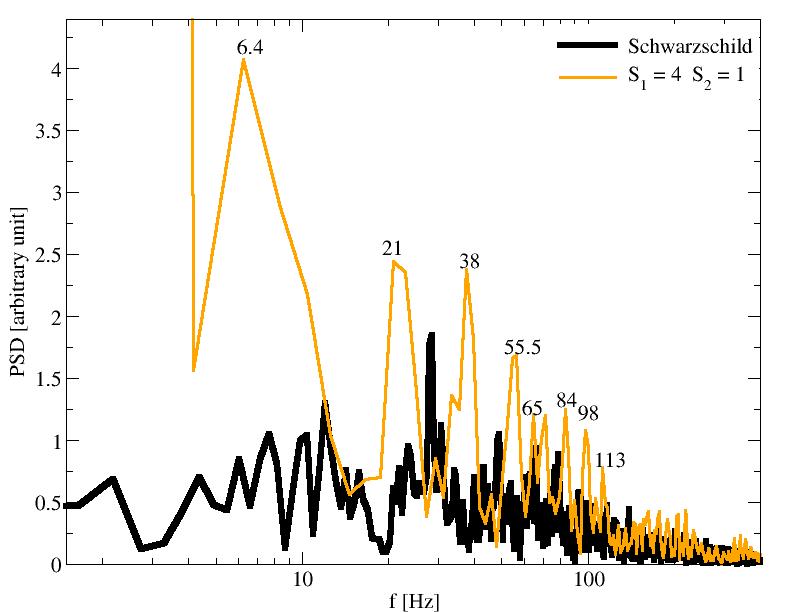}
\includegraphics[width=8.0cm,height=7.0cm]{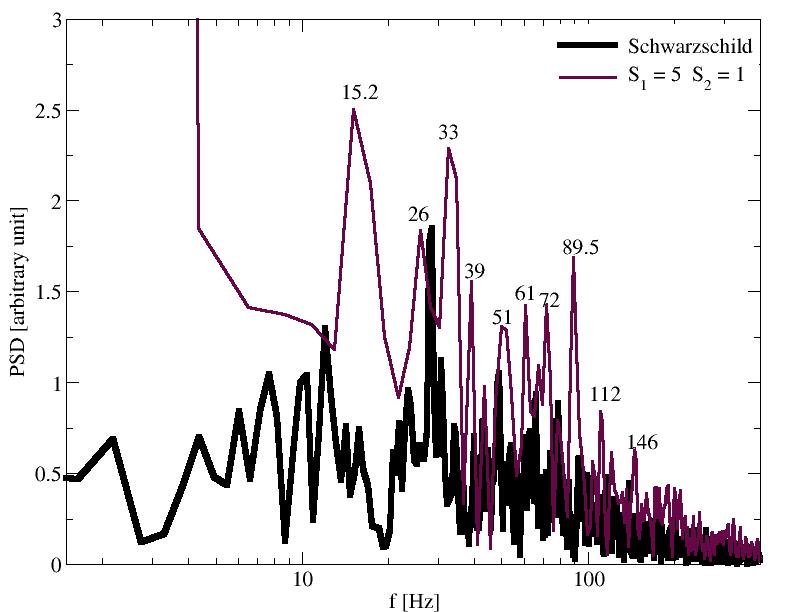}
\caption{The same analysis as in Fig.\ref{QPOs1} is performed, but this time the PSD results are shown for varying $S_1$ while keeping $S_2$ fixed.  Similar quasi-periodic behavior is also observed in this case.
}\label{QPOs2}
\end{figure*}

To test the physical detectability of the Lee-Wick BH, it is essential to investigate the QPO modes that are trapped and excited within the shock cone mechanism forming around the BH. A comparative analysis between the Lee-Wick and Schwarzschild spacetimes allows us to identify both their similarities and differences, revealing how observational signals could carry the imprints of Lee-Wick gravity. Based on this motivation and considering that the strong oscillations demonstrated in Fig. (\ref{massAcc}) arise from the variability of the mass accretion rate, we performed a PSD analysis of the accretion rate. Fig. (\ref{QPOs1}) and Fig. (\ref{QPOs2}) present the PSD computed using the mass-accretion rate at $r = 2.3M$, for various combinations of parameters $(S_{1}, S_{2})$ of the Schwarzschild and Lee-Wick models. To express the resulting frequencies in units of Hertz, we set the BH mass at $M = 10 M_{\odot}$. After confirming that the QPO modes that arise in the PSD are global in nature, rather than the product of local instabilities, the analysis was carried out using only the accretion rate at $r = 2.3M$. Numerical experiments performed with the accretion rate at $r = 6.11M$ produced identical QPO frequencies, confirming the robustness of the signal. In Fig. (\ref{QPOs1}), the PSD derived from the strongly oscillatory accretion rate at $r = 2.3M$ displays the QPO peaks around the Schwarzschild and Lee-Wick BHs for constant $S_{1}$ and varying $S_{2}$. The upper left panel corresponds to the Schwarzschild case, showing a single broad LFQPO peak, which represents the typical post-shock oscillation observed in classical BHL accretion. When the Lee-Wick correction term modifies the spacetime geometry, additional sharp peaks appear in the PSD. The frequencies of these peaks increase systematically with larger $S_{2}$, causing the QPOs to shift from the LFQPO to the high-frequency QPO (HFQPO) regime. This result, consistent with theoretical expectations, confirms that the strong oscillatory term in the Lee-Wick metric amplifies the curvature-driven instabilities within the shock cone.

For low values of $S_{2}$ ($S_{2}=1$), the observed peaks exhibit an approximate 3:2 ratio, likely arising from nonlinear resonances between the radial and azimuthal oscillation modes. As $S_{2}$ increases, these ratios evolve toward 2:1 or 3:1, revealing higher harmonics and complex mode coupling indicative of stronger Lee-Wick effects. Moreover, similar mode ratios are observed in different combinations of $(S_{1},S_{2})$, consistent with observational data. The rich set of peaks obtained in each model arises from the superposition of fundamental oscillation modes, leading to new resonant frequencies. Consequently, when $S_{1}$ is kept constant, increasing $S_{2}$ increases the probability that the observable QPO frequencies correspond to HFQPOs.

In Fig. (\ref{QPOs1}), it was shown that the increase in the parameter $S_2$ in the PSD analysis causes the characteristic frequencies to evolve toward the HFQPO regime. Consequently, the mode ratios emerging from the simulations are observed to be most prominently $3:2$, with additional branches $2:1$ and $4:3$. In particular, the numerically detected $3:2$ HFQPO pairs correspond well to those observed in stellar-mass BH systems, such as $GRO J1655--40$ ($300/450$), $XTE J1550-564$ ($184/276$), and $GRS 1915+105$ ($41/67$) \cite{Kluzniak2005,Remillard2006,Belloni2014,Sreehari2020}. These observational findings confirm the presence of stable $3:2$ frequency ratios in real astrophysical systems. However, the appearance of $2:1$ ratios corresponds to transient states typically observed during state transitions in X-ray binaries. Consequently, the numerically calculated QPO ratios around the Lee-Wick BH, especially for parameter sets $S_1 = 6$, $S_2 = 1$ (clear $3:24$) and $S_1 = 6$, $S_2 = 6$ (\( \approx 3:2\) with a nearby $2:1$ branch), are in strong agreement with the observed HFQPO ratios in BH binaries. The presence of additional $4:3$ branches reflects intermediate resonant behavior, which has also been discussed in relativistic resonance interpretations of QPO phenomena.

In Fig. (\ref{QPOs2}), the PSD analysis is repeated, keeping $S_{2}$ fixed and varying $S_{1}$. In this case, the quasi-periodic nature of the signal persists across all models, but the dominant peaks gradually shift toward lower frequencies, indicating a transition toward LFQPOs. This behavior is attributed to the strong exponential damping term, which deepens the effective potential and lengthens the characteristic oscillation period of the accretion flow. Due to nonlinear coupling among oscillation modes, the PSDs exhibit a rich structure, making the 3:2 and 2:1 frequency ratios highly likely to be observed around the Lee-Wick BH. As seen in Fig. (\ref{QPOs2}), when $S_{2}$ is fixed, increasing $S_{1}$ not only modifies the frequency positions but also enhances the amplitude of the PSD peaks. Thus, the detectability of such peaks in observational data becomes more probable.

For the models presented in Fig. (\ref{QPOs2}), the mode ratios obtained from numerical calculations are highly consistent with the observed HFQPO frequency pairs in X-ray binary systems. For the case of $S_1 = 1$ and $S_2 = 1$, the dominant peaks and their ratios are found as $48/34 \approx 1.41$ ($\approx 3:2$) and $99/48 \approx 2.06$ ($\approx 2:1$). At $S_1 = 2$, $S_2 = 1$, the numerically obtained ratios are $32/23 \approx 1.39$ ($\approx 4:3$) and $47/23 \approx 2.04$ ($\approx 2:1$), indicating both resonant and overtone-type harmonics. In the case of $S_1 = 4$, $S_2 = 1$, the ratios $38/21 \approx 1.81$ ($9:5 \approx 2:1$) and $55.5/38 \approx 1.46$ ($\approx 3:2$) are obtained. Finally, for $S_1 = 5$, $S_2 = 1$, the observed ratios are $33/26 \approx 1.27$ ($\approx 4:3$) and $72/51 \approx 1.41$ ($\approx 3:2$). These numerically observed mode pairs are in strong agreement with the 3:2 HFQPO pairs detected in BH X-ray binaries such as $GRO J1655-40$ ($300 / 450 $), $XTE J1550-564$ ($184/ 276$), and $GRS 1915+105$ ($41/ 67$) \cite{Kluzniak2005,Remillard2006,Belloni2014,Sreehari2020}. In addition, the $2:1$ and $4:3$ branches found in numerical simulations correspond well to transient resonant behaviors observed in systems such as $XTE J1859+226$ \cite{Homan2003} and $H1743-322$ \cite{Torok2005}.

Compared to the Schwarzschild case, the Lee-Wick models produce a broader and richer PSD structure. Around the Schwarzschild BH, the QPO frequencies exhibit relatively low amplitudes dominated by LFQPOs. In contrast, the Lee-Wick parameters $S_{1}$ and $S_{2}$ reshape the spacetime geometry, amplifying the turbulence within the shock cone and enhancing the observability of strong, high-amplitude QPO peaks. Increasing $S_{1}$ strengthens the gravitational confinement and slightly lowers the oscillation frequencies, whereas increasing $S_{2}$ intensifies the oscillatory curvature component, generating more complex modulations and noticeable frequency up-shifts. These results demonstrate that the Lee-Wick corrections produce observable differences in the PSD signatures, offering a promising avenue for identifying potential quantum-gravity imprints in accreting BH systems. 

\subsection{\text{Block-2}: Strong Lee-Wick Regime}\label{B2} 

In the \text{Block-2} configurations presented in Tab. (\ref{Inital_Con_1}), the Lee-Wick BH exhibits strong spacetime corrections. In this regime, both the exponential term governed by $S_{1}$ and the oscillatory term determined by $S_{2}$ become significantly more pronounced in shaping the metric function. The parameters $S_{1}$ and $S_{2}$ attain values that amplify the higher contributions from the derivative, thereby leading to substantial deviations from the Schwarzschild solution and, in some cases, to the formation of multiple horizons. The oscillatory component associated with $S_{2}$ introduces spatially varying corrections to the gravitational potential, reflecting the characteristic behavior of complex mass poles in the Lee-Wick formulation. As a result, the curvature profile of spacetime may develop nontrivial features that encode the influence of quantum-scale modifications of gravity. The parameter values used in \text{Block-2}, therefore, enable us to explore how the fundamental structure of spacetime is altered in the strong Lee-Wick regime. As discussed in detail in the following, the numerical modeling of this regime reveals that strong Lee-Wick corrections modify the overall morphology of the accretion flow, alter the formation of the bow shock, and significantly influence the QPOs that arise within or near the shock region.

\subsubsection{Morphology of Accretion Flow and Shock-Cone Dynamics}\label{Morph_B2} 

\begin{figure*}
\centering 
\includegraphics[width=7.5cm,height=7.5cm]{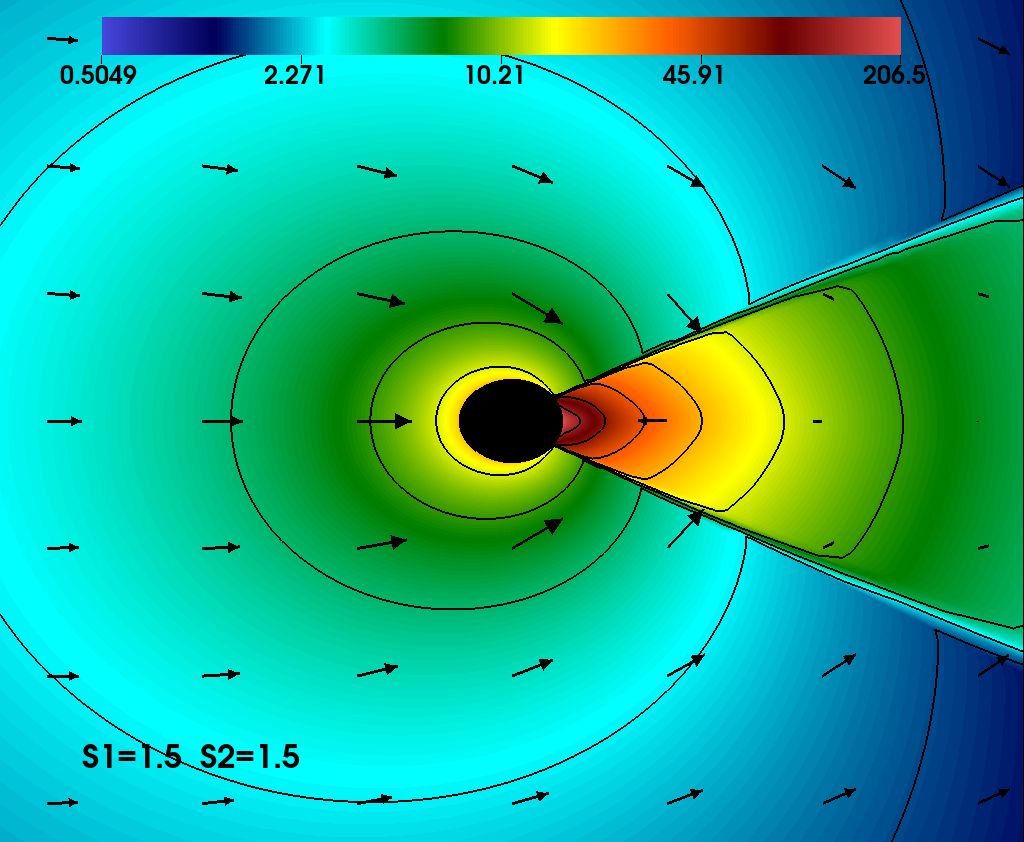} 
\includegraphics[width=7.5cm,height=7.5cm]{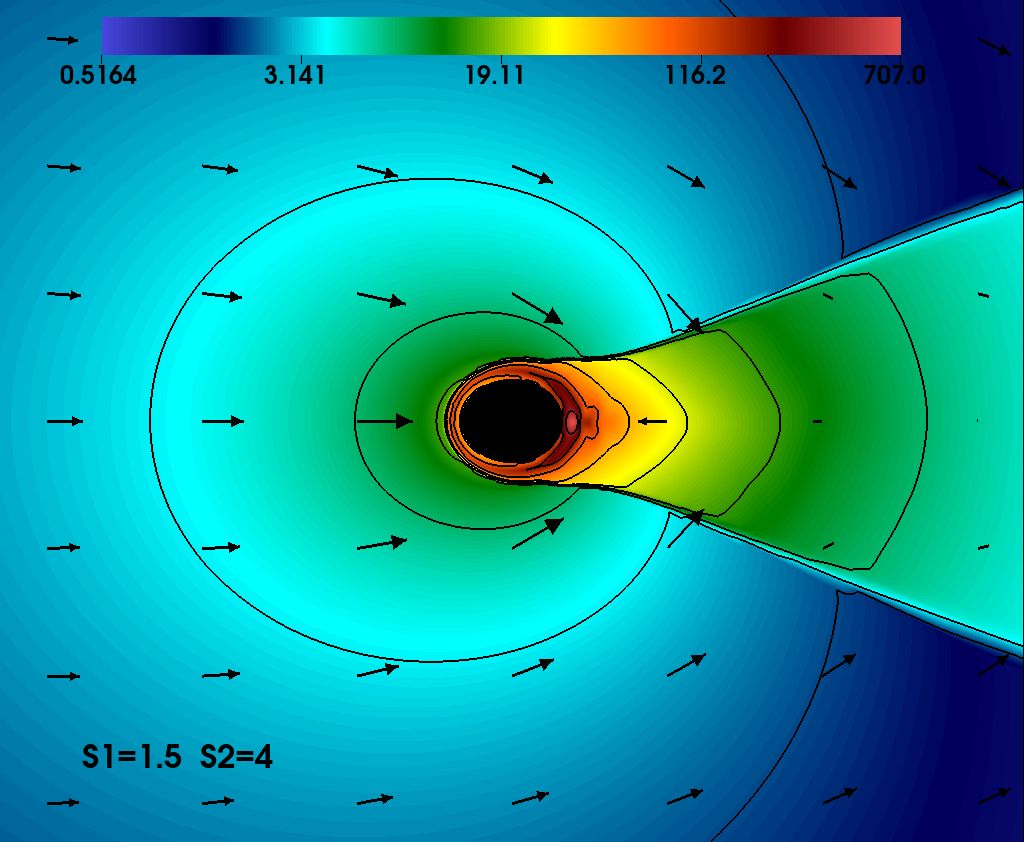}\\
\includegraphics[width=7.5cm,height=7.5cm]{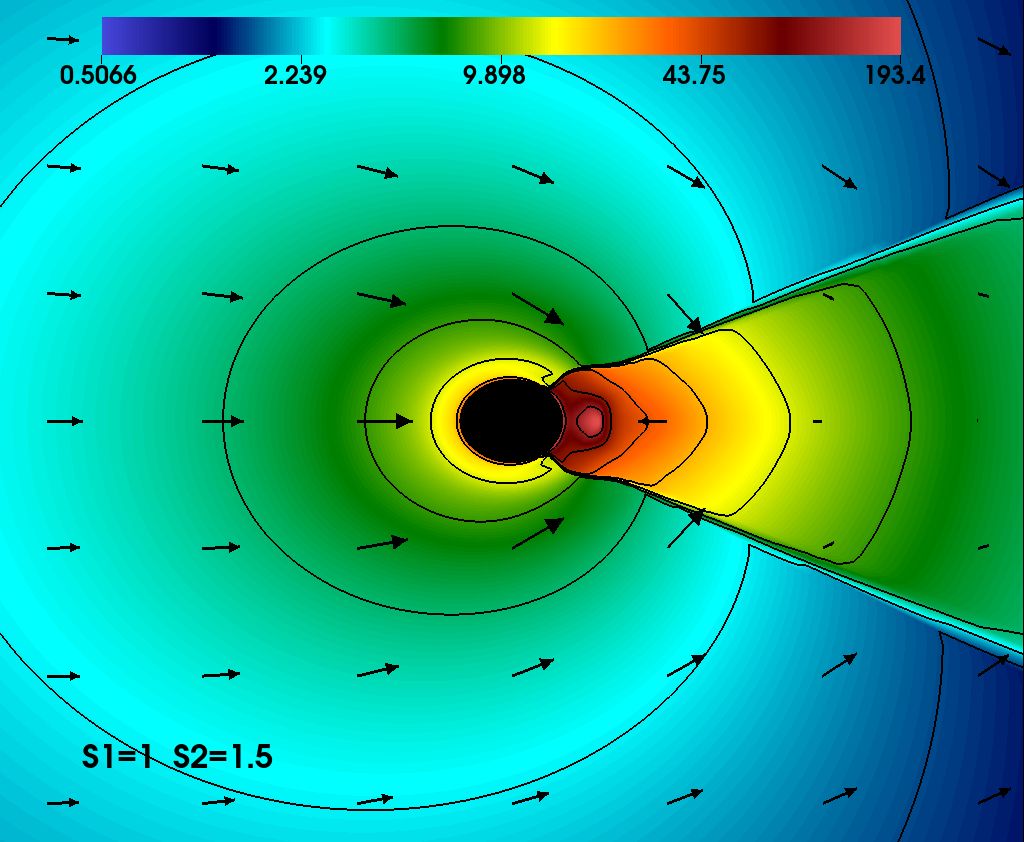} 
\includegraphics[width=7.5cm,height=7.5cm]{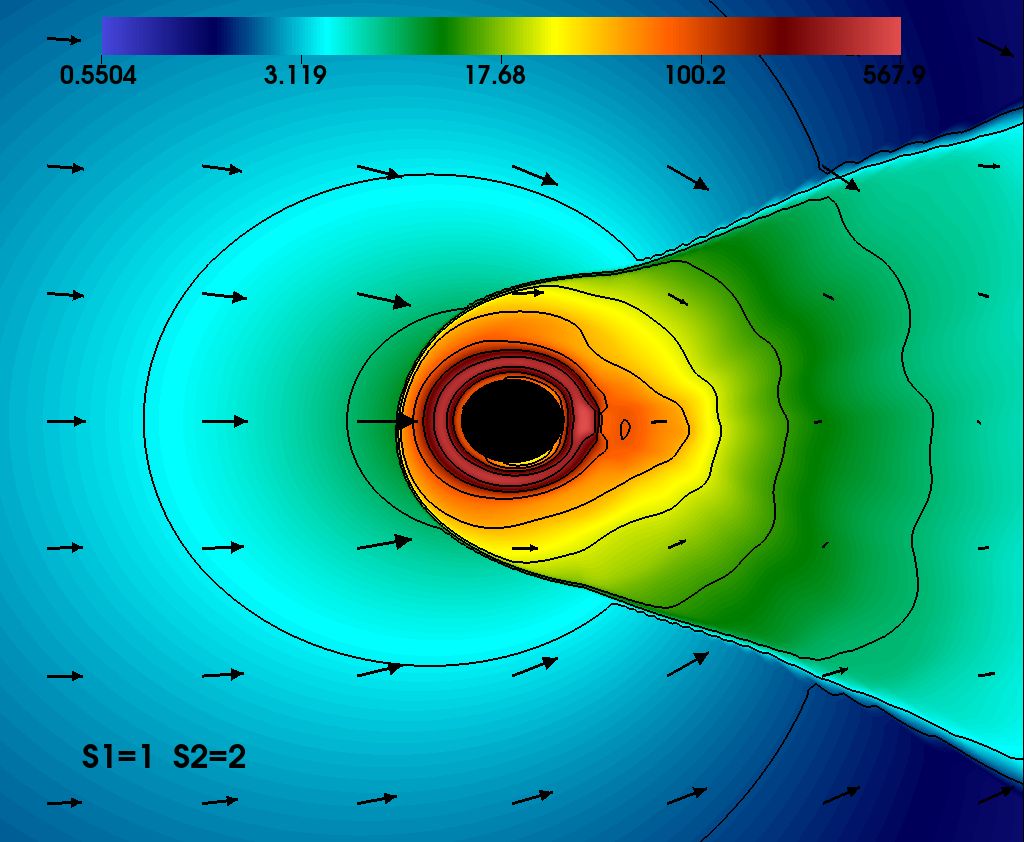}
\caption{Same as Fig. (\ref{color_block1}), but this time the rest-mass density is shown, particularly near the BH, for the models in \text{Block-2} listed in Tab. (\ref{Inital_Con_1}). It can be seen that, depending on the model, the shock cone forms around the BH and consequently produces a bow shock or strong instabilities.
}\label{color_block2}
\end{figure*}

In order to reveal the effect of the parameter $S_{2}$, which represents the oscillatory component of $f(r)$, on the shock cone structure formed around the non-rotating Lee-Wick BH, the numerical results presented in Fig. (\ref{color_block2}) have been analyzed. In Fig. (\ref{color_block2}), using the initial conditions given in \text{Block-2} of Tab. (\ref{Inital_Con_1}), the rest mass density color and contour plots, together with the velocity vector field, are presented to illustrate the morphology of the shock cone produced by BHL accretion and to demonstrate the influence of parameters $S_{1}$ and $S_{2}$ on the cone structure. In the upper-left snapshot, corresponding to $S_{1}=1.5$ and $S_{2}=1.5$, it is observed that both the exponential ($S_{1}$) and oscillatory ($S_{2}$) terms moderately affect the morphology of the shock structure. In this case, as a result of BHL accretion on the downstream side, a well-defined and symmetric shock cone similar to those of \text{Block-1} (Fig.\ref{color_block1}) is formed. Since $S_{1}$ is relatively large, the exponential term becomes dominant, and thus the Lee-Wick metric behaves similarly to the Schwarzschild case, producing a stable and narrow shock cone. In this regime, the spacetime curvature acts as a strong focusing potential with little oscillatory distortion.

In the upper right panel of Fig. (\ref{color_block2}), the cases $S_{1}=1.5$ and $S_{2}=4$ are modeled. Due to the larger value of $S_{2}$, this configuration exhibits rapid fluctuations in curvature. As seen in Fig.\ref{color_block2}, the cone becomes wider and encloses the BH, forming a bow-shock-like structure. The resulting pattern shows asymmetric density contours and visible instabilities near the cone’s apex. This behavior arises because the oscillatory curvature term governed by $S_{2}$ generates varying local gravitational gradients, leading the flow to overshoot and produce quasi-periodic pressure waves.

In the lower-left panel of Fig. (\ref{color_block2}), the cases $S_{1}=1$ and $S_{2}=1.5$ are considered, where the lower value $S_{1}$ corresponds to a weaker exponential damping. Consequently, the Lee-Wick corrections remain strong even far from the horizon. The cone opening angle increases, and the density gradient becomes flatter, indicating a weakening of the gravitational confinement. However, since $S_{2}$ is moderate, the oscillations remain mild and the flow is mostly stable, representing a transitional regime between Schwarzschild-like and oscillatory dynamics. In this case, a region of enhanced turbulence appears near the BH horizon.

In the lower-right panel of Fig. (\ref{color_block2}), the numerical results for $S_{1}=1$ and $S_{2}=2$, a configuration that favors strong Lee-Wick effects, are shown. Here, the exponential suppression is weak while the oscillatory curvature is prominent. This setup produces a dense, thick shock region that almost completely surrounds the BH, forming a bow shock. The formation of this bow shock arises from the collision of supersonically flowing matter with the high-pressure cavity in the downstream region. For larger $S_{2}$ values, the enhanced oscillatory behavior modifies the effective potential around the Lee-Wick BH, increasing local compressions. This leads to nonlinear coupling between radial and azimuthal modes within the bow shock, giving rise to strong quasi-periodic oscillatory behavior. The development of these shock cones and bow shocks is in complete agreement with the theoretical analyzes discussed above and with the effective potential, force profile, and ISCOs results presented in Fig.  (\ref{fig_Veff}), Fig. (\ref{fig_AngMom}), Fig. (\ref{fig_Energy}), and Fig. (\ref{fig_ISCO}).

\begin{figure*}
\centering 
\includegraphics[width=15.0cm,height=11.5cm]{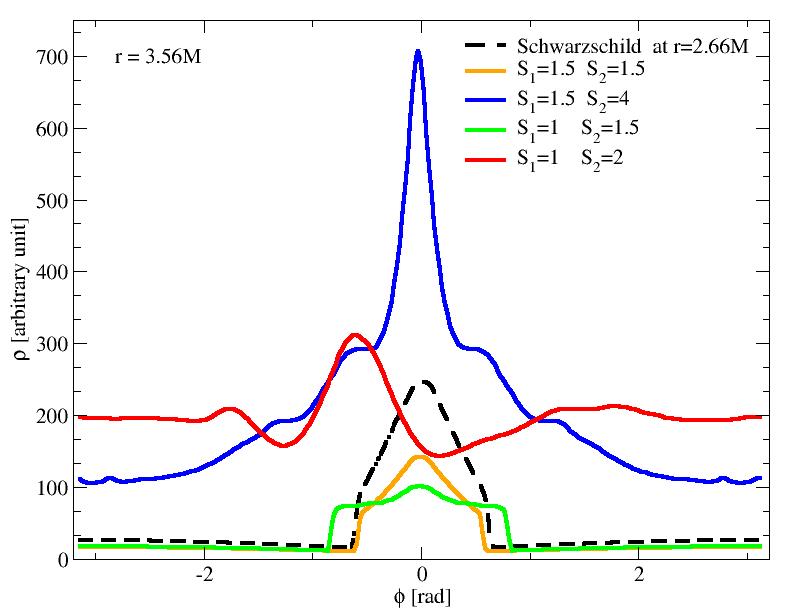} \\
\includegraphics[width=15.0cm,height=11.5cm]{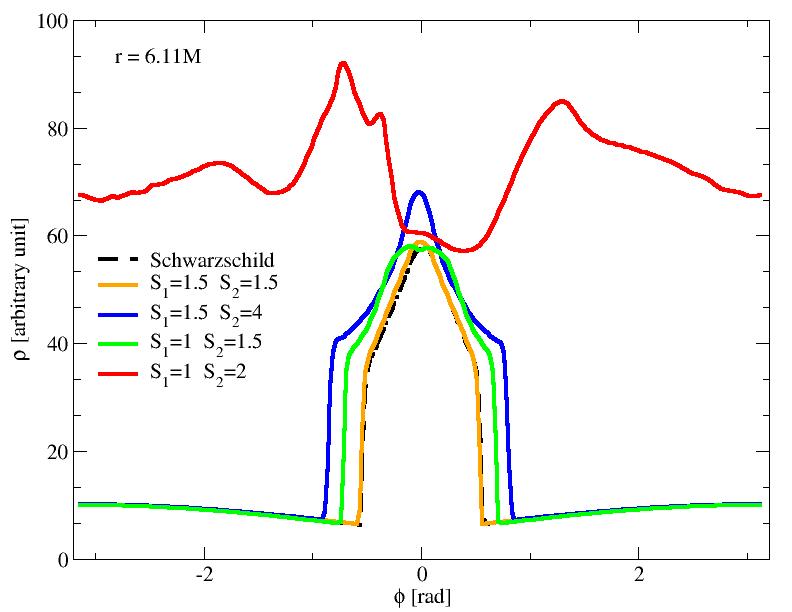} 
\caption{Similar to Fig. (\ref{azimutal_den}), the azimuthal variation of the rest-mass density at two different radial locations is shown for the \text{Block-2} models listed in Tab. (\ref{Inital_Con_1}).  In the upper panel, the variation of the rest-mass density at $r = 3.56M$ is presented for different $S_1$ and $S_2$ values and compared with the Schwarzschild case.  In the lower panel, the azimuthal behavior of the rest-mass density around the ISCOs, at $r = 6.11M$, is displayed.  It is clearly observed that, especially at $r = 3.56M$, the region closest to the horizon, the behavior of the rest mass density in Lee-Wick gravity significantly deviates from that of the Schwarzschild spacetime. 
}\label{azimutal_den_Block_2}
\end{figure*}

Analyzing the azimuthal variation of the rest-mass density is crucial for understanding the dynamical structure of the physical system that forms in the strong gravitational field and near the ISCOs region, as well as for revealing the influence of the Lee-Wick parameters on spacetime geometry. Therefore, in Fig. (\ref{azimutal_den_Block_2}), the azimuthal variation of the rest-mass density is presented. The top panel shows the behavior in the strong gravitational field at $r = 3.56M$, while the bottom panel illustrates the distribution near the ISCOs at $r = 6.11M$. In each panel, using the \text{Block-2} initial conditions listed in Tab. (\ref{Inital_Con_1}), the wild behaviors arising from the spacetime modifications induced by the Lee-Wick parameters are demonstrated. As observed in $r = 3.56M$, almost all models exhibit significant deviations compared to the Schwarzschild case. In particular, in the two models with higher values of oscillatory parameters ($S_{2} = 2$ and $S_{2} = 4$), pronounced and asymmetric peaks in density appear, revealing the formation of strong shock waves due to enhanced curvature oscillations. These strong fluctuations cause the accreting matter near the BH horizon to compress sharply, forming a bow-shaped shock cone surrounding the BH. In contrast, in cases with moderate or smaller $S_{2}$ values ($S_{2} = 1.5$), the density profile becomes smoother and a weaker shock wave is produced. This behavior originates from the exponential damping determined by $S_{1}$, which leads to a more stable and symmetric accumulation of matter around the BH.

As seen in the lower panel of Fig. (\ref{azimutal_den_Block_2}) at $r = 6.11M$, corresponding to the ISCOs region, the influence of the Lee-Wick corrections is significantly weakened. Consequently, almost all models converge toward the Schwarzschild solution, although the maximum value of the rest-mass density remains relatively lower compared to the Schwarzschild case. However, the model with the highest oscillatory parameter value ($S_{1}=1$, $S_{2}=2$) still shows a clear deviation from Schwarzschild behavior. This indicates that increasing $S_{2}$ not only alters the near-horizon structure but also significantly modifies the morphology of the physical flow even in slightly distant regions. These numerical results are in complete agreement with the theoretical predictions discussed earlier in the paper. The increasing $S_{2}$ enhances the oscillatory mode of curvature and modifies the effective potential, resulting in stronger local gravitational gradients, nonlinear coupling between modes, and the formation of intense shock structures near the BH.

\begin{figure*}
\centering 
\includegraphics[width=15.0cm,height=12.0cm]{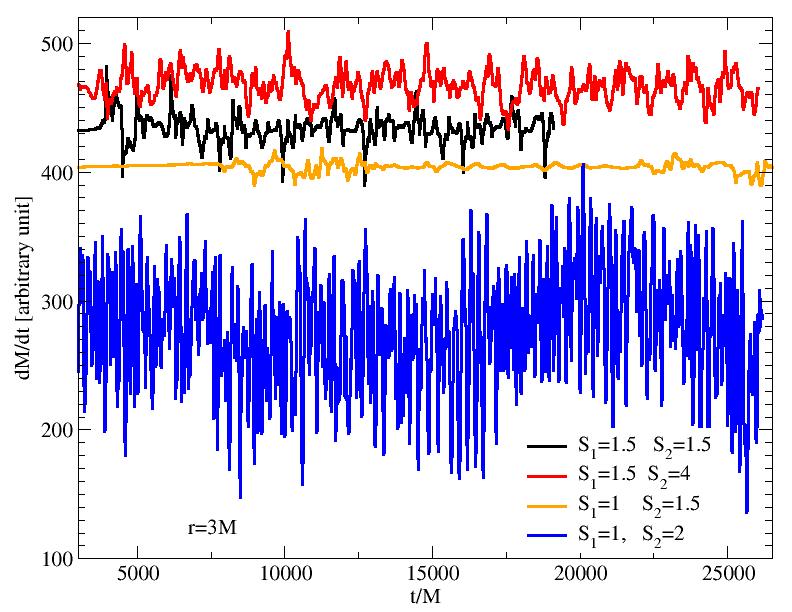} 
\caption{For the models given in \text{Block-2} of Tab. (\ref{Inital_Con_1}), the time evolution of the mass accretion rate is presented. The figure illustrates how the oscillation strength of the shock cone and bow shock varies with different values of the Lee-Wick gravity parameters \(S_{1}\) and \(S_{2}\). Increasing the parameter \(S_{2}\) enhances the oscillatory behavior of the accretion flow, leading to stronger temporal fluctuations and quasi-periodic variations in the mass accretion rate.
}\label{massAcc_Block2}
\end{figure*}

For the \text{Block-2} models given in Tab. (\ref{Inital_Con_1}), the time evolution of the mass accretion rate calculated at $r = 3M$ is presented in Fig. (\ref{massAcc_Block2}). As clearly seen in each model, different levels of instabilities are produced. The most stable configuration, characterized by the smallest oscillation amplitude, corresponds to the case with the lowest $S_{2}$ value and the strongest exponential damping term. Its fluctuations are low in amplitude and quasi-regular, consistent with a spacetime metric closer to the Schwarzschild geometry and with weaker post-shock turbulence. In Fig. (\ref{massAcc_Block2}), when $S_{1}$ is fixed and $S_{2}$ increases, the signal becomes noticeably noisier and broadband, with significantly larger oscillation amplitudes compared to the other models. The enhanced oscillatory component of the lapse function $f(r)$ drives stronger curvature ripples, which feed variability into the shock cone/bow-shock system and the downstream cavity. This leads to the excitation of strong global oscillation modes.

When the exponential term is suppressed (i.e., smaller $S_{1}$), both the mean accretion rate and its temporal variation increase, since the effective gravitational force becomes more attractive and the infalling matter is compressed more efficiently before forming a shock. As a result, larger values $S_{2}$ amplify the oscillatory nature of the metric and destabilize the post-shock flow. At the same time, as shown in the theoretical analyzes presented earlier, increasing $S_{1}$ or $S_{2}$ deepens the effective potential, thereby enhancing the gravitational attraction. As also predicted by theory, stronger compression near the BH leads to shifts in the ISCOs position, altogether yielding the stronger, less-regular variability observed in the high $S_{2}$ runs.  

The oscillatory curvature creates rapidly varying local gravitational gradients and, as a result, strong instabilities appear, as evident in Fig. (\ref{massAcc_Block2}). The accretion flow repeatedly overshoots, reflects within the high-pressure cavity, and drives standing and advected shock waves, leading to large-amplitude and irregular fluctuations in the mass accretion rate. This mechanism also explains the different shock morphologies observed in Fig. (\ref{color_block2}). In all model configurations, the generated instabilities represent key features that support the formation of QPOs. In particular, the strong oscillations observed in the high cases $S_{2}$ correspond to quasi-periodic variations, confirming the emergence of nonlinear oscillatory behavior consistent with the QPO formation scenario discussed in Section \ref{QPOs_B2}.

\subsubsection{Oscillatory Accretion and Quasi-Periodic Signatures }\label{QPOs_B2} 

\begin{figure*}
\centering 
\includegraphics[width=8.0cm,height=7.0cm]{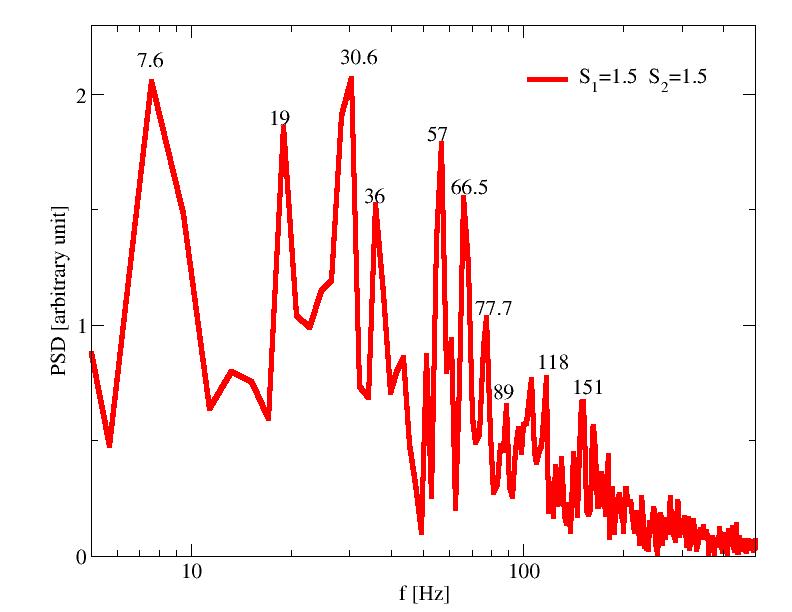} 
\includegraphics[width=8.0cm,height=7.0cm]{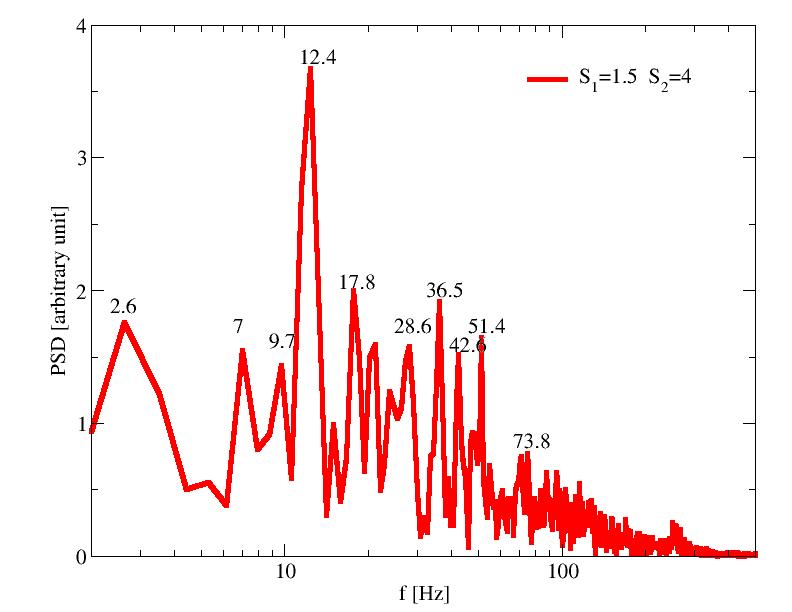} 
\includegraphics[width=8.0cm,height=7.0cm]{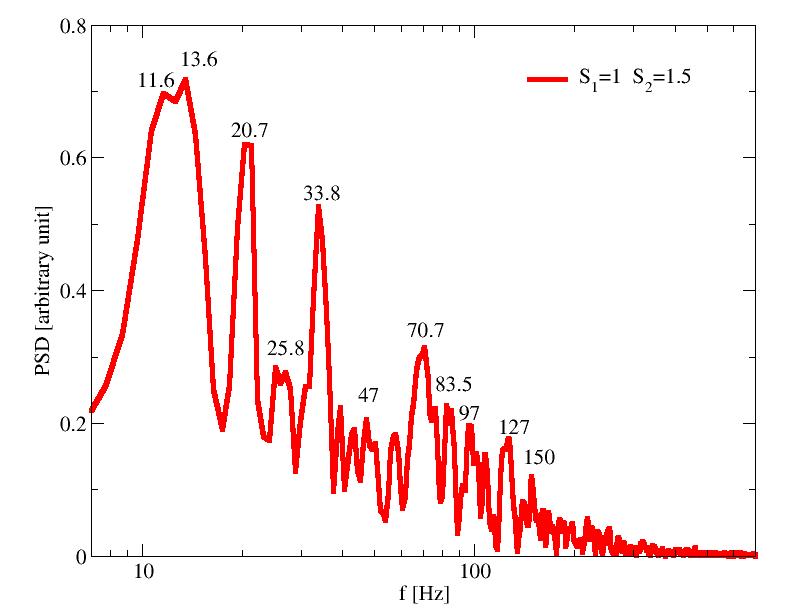}
\includegraphics[width=8.0cm,height=7.0cm]{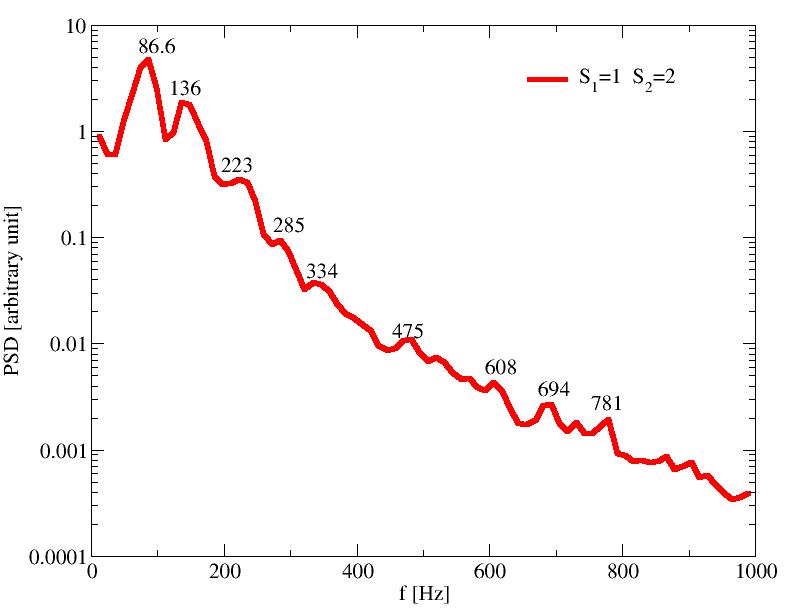}
\caption{PSD analysis of the mass accretion rate for the initial conditions given in the \text{Block-2} configuration, showing the dependence on the Lee-Wick parameters. For each model, a rich set of oscillation frequencies is observed, clearly demonstrating the influence of the Lee-Wick parameters on the resulting QPO frequencies.  Increasing the parameter \(S_{2}\) shifts the dominant frequencies toward the HFQPO regime, indicating stronger nonlinear coupling between the radial and azimuthal oscillation modes within the accretion disk. 
}\label{QPOs_block2}
\end{figure*}

As seen in Fig. (\ref{massAcc_Block2}), the mass accretion rate exhibits strong oscillatory behavior depending on the parameters $S_{1}$ and $S_{2}$, reflecting the dynamical influence of the shock cone or the bow-shock structure formed around the BH. For each \text{Block-2} model with different $S_{1}$ and $S_{2}$ values, PSD analyzes are performed using the mass accretion rate data shown in Fig.(\ref{massAcc_Block2}). As clearly observed, each configuration displays a rich harmonic structure, and the resulting QPO analyzes are presented in Fig.(\ref{QPOs_block2}). In Fig.(\ref{QPOs_block2}), for the cases of $S_{1}=1.5$ and $S_{2}=1.5$, the PSD analysis reveals that the QPO frequencies appear in the range of $7-70$ Hz. The frequency ratios are approximately $2:1$ $\approx 19/10$, $3:1$ $\approx 30.6/10$, and $4:1$ $\approx 36/10$, corresponding to weakly coupled LFQPOs in a spacetime similar to Schwarzschild. When $S_{2}$ increases to $4$, the dominant frequency shifts to about $12$ Hz, with several overtones that extend to $\sim70$ Hz. The characteristic ratios in this model are $3:2$ $\approx 17.8/12.4$, $3:1$ $\approx 36.5/12.4$, $4:1$ $\approx 51.4/12.4$, and $6:1$ $\approx 73.8/12.4$, indicating the presence of a strong nonlinear coupling between oscillation modes.  

For the configurations $S_{1}=1$ and $S_{2}=1.5$, the PSD analysis shows slightly higher fundamental modes in the range of $11-20$ Hz, suggesting a moderate oscillatory modulation of the accretion flow. The observed frequency ratios are $3:2$ $\approx 20.7/13.6$, $2:1$ $\approx 25.8/13.6$, and $5:2$ $\approx 33.8/13.6$, revealing harmonically related overtones. However, the most striking behavior appears in the cases of $S_{1}=1$ and $S_{2}=2$, where the PSD displays peaks extending up to nearly $800$ Hz, with dominant features at $86$, $136$, $223$, $285$, and $324$ Hz. These yield characteristic ratios of $3:2$, $5:2$, $3:1$, and $4:1$, which are typical of HFQPO.  

These results clearly show that the parameter $S_{1}$ controls the exponential damping of the quantum corrections, enhancing the stability of accretion and favoring the appearance of low-frequency modes. In contrast, increasing $S_{2}$ strengthens the oscillatory component of the metric, amplifies nonlinear curvature coupling, and shifts the resulting oscillation frequencies toward the HFQPO regime. The transition from LFQPO to HFQPO in the simulations corresponds to the inward shift of the effective potential and ISCOs as $S_{2}$ increases, leading to faster epicyclic oscillations consistent with theoretical predictions of the Lee-Wick spacetime.  

Finally, the QPOs obtained from the numerical simulations confirm that the Lee-Wick parameters naturally allow for the coexistence of multiple QPO regimes. The LFQPOs observed in low$S_{2}$ cases, typically between $7$ and $30$ Hz, are comparable to those detected in black-hole X-ray binary systems. In contrast, high-$S_{2}$ models produce QPOs in the range of $80$ to $300$ Hz, consistent with observed HFQPOs, such as $\sim67$ Hz in GRS 1915+105, $\sim184/276$ Hz and $\sim300/450$ Hz $3: 2$ pairs in GRO J1655-40 and XTE J1550-564 \cite{Kluzniak2005,{Remillard2006},Belloni2014,Sreehari2020}. In particular, the $136/86 \approx 3:2$ ratio obtained for the $S_{1}=1$, $S_{2}=2$ model closely matches the observational data from GRS 1915+105, confirming the physical relevance of the simulated QPO frequencies to real astrophysical systems.

\section{Observational Implications of \text{Block-1} and \text{Block-2} Accretion Regimes}\label{compare_obs} 

According to the numerical results presented in Section \ref{NumSim}, the morphology of the accretion flow and its time-dependent variations around the nonrotating Lee-Wick BH are strongly dependent on the two coupling parameters, \(S_{1}\) and \(S_{2}\), which characterize the space-time metric of this BH. While \(S_{1}\) represents the exponential component of the metric function \(f(r)\), \(S_{2}\) corresponds to its oscillatory component. In the parameter region defined as \text{Block-1} in Tab. (\ref{Inital_Con_1}), where the corrections are weak (\(S_{1} \geq 2\) and \(S_{2} \leq 1.5\)), the exponential damping dominates, causing the spacetime geometry to behave almost identically to that of the Schwarzschild solution. Consequently, the BHL accretion pattern forms a smooth, narrow, and quasi-steady-state shock structure. The corresponding temporal evolution of the mass-accretion rate exhibits mild periodic modulations with low amplitude, producing LFQPOs-like features with harmonics around a few tens of hertz. These results are consistent with type-C LFQPOs observed in Galactic black-hole binaries such as XTE J1550-564 and GRS 1915+105. Therefore, it can be inferred that the QPO signals detected from these sources likely originate from stable shock oscillations and weak nonlinear coupling effects within the accretion flow in a spacetime where the Lee-Wick corrections are negligible.

In contrast, when the Lee-Wick metric is characterized by lower values of \(S_{1}\) and higher values of \(S_{2}\), strong oscillations arise, leading to rapid curvature fluctuations and even reversals of the gravitational gradient. As a result, the accretion flow becomes highly dynamical, forming broad and asymmetric shock cones as well as distinct bow-shock morphologies. The generated bow shock entirely encloses the BH, enhancing the strength of the oscillatory behavior. The temporal evolution of the accretion rate in this regime reveals strong, short-time-scale quasi-periodic modulations. The PSD analyzes demonstrate that dominant peaks occur within the $10-800$ Hz range. These peaks exhibit frequency ratios such as $3:2$, $2:1$, $3:1$, and $5:2$, which arise due to the strong nonlinear coupling between the radial and azimuthal oscillation modes in the region of intense dynamical activity, namely, within the shock cone and bow-shock zones. The numerical results clearly indicate that as \(S_{2}\) increases, the characteristic QPOs systematically shift toward the HFQPO regime. These findings are in full agreement with the theoretical results presented in Section \ref{oscillations}, where it was shown that the oscillatory curvature enhances the radial epicyclic frequency and causes the ISCOs to move inward. In physical terms, a larger \(S_{2}\) increases the oscillatory component of the gravitational potential, thereby amplifying the nonlinear mode coupling and pushing the disk oscillations into the relativistic regime.

When the agreement between the numerical results and observations is considered, the frequency ranges obtained from the simulations show strong consistency with those observed in microquasars and X-ray binary systems. In the \text{Block-1} regime, the LFQPOs identified in the $5-30$ Hz range are consistent with the type-B/C QPOs observed in sources such as XTE J1550-564 \cite{Liu2022etal}, GX 339-4 \cite{Nikolaos2020} and GRS 1915 + 105 \cite{Zhang2020etal}. In contrast, HFQPOs found in the \text{Block-2} regime, within the $80-300$ Hz range, coincide with the $67$ Hz oscillation of GRS 1915+105 , the $184/276$ Hz pair of XTE J1550-564 \cite{Varniere2018}  and the pair of $300/450$ Hz detected in GRO J1655-40 \cite{Mendez2013etal}. In particular, the $3:2$ frequency ratio observed in the PSD analysis for the cases \(S_{1}=1\) and \(S_{2}=2\) exhibits a strong parallel to the resonance ratios reported in these observational sources. Moreover, the amplitude growth and spectral broadening observed in the numerical simulations for large \(S_{2}\) values are remarkably similar to the irregular variability and turbulence-induced transitions detected during state changes of transient BH binary systems.

The shift of the numerically extracted QPOs can also be understood from the theoretical analysis of the effective potential and ISCOs structure presented in Section \ref{part_dynam}. As shown in the paper, increasing \(S_{1}\) deepens the minimum of the potential through exponential suppression, while increasing \(S_{2}\) superposes an oscillatory term on the potential. This addition modulates the curvature and splits the potential well into sub-minima. The strong interplay between these two terms naturally leads to the formation of multiple resonance cavities, within which matter undergoes alternating phases of compression and rarefaction. As a result, both LFQPOs and HFQPOs are effectively excited. This behavior closely parallels discrete peaks and broadband noise observed in systems such as H1743-322 \cite{Remillard2006etal} and 4U 1630-47 \cite{Chopra2025}, where the presence of multiple QPO peaks has been confirmed observationally. Therefore, the numerical simulations performed in this study provide a physically motivated mechanism through which higher-derivative gravitational corrections, parameterized by \((S_{1}, S_{2})\), can leave measurable imprints on the temporal spectra of accreting BHs.

From the results obtained in the numerical simulations, two distinct observational implications can be identified. First, according to the results derived within the Lee-Wick framework, the QPO frequencies formed around nonrotating BHs depend not only on the mass of the BH but also on the parameters \(S_{1}\) and \(S_{2}\). This introduces new degrees of freedom that could be constrained by future $X-$ray timing observations. Next-generation missions such as eXTP \cite{Zhang2025etal}, STROBE-X \cite{Froning2024etal}, and Athena \cite{Hauf2011etal} may be able to test whether the observed frequency ratios and their energy dependencies correspond to the parameter dependence on \(S_{1}\) and \(S_{2}\) discussed in detail here. Second, the morphological structure of the bow shock that emerges at large \(S_{2}\) values may manifest itself as asymmetric brightness features or variable absorption patterns in Event Horizon Telescope scale \cite{EHT1} images of accretion flows, offering a potential multiwavelength diagnostic of quantum-gravity corrections. Furthermore, the oscillatory curvature and enhanced compression that develop particularly near the horizon could affect the jet-launching conditions, potentially linking the Lee-Wick parameter \(S_{2}\) to episodic radio flaring similar to that observed in GRS 1915+105 \cite{Motta2021etal}. Lastly, the Lee-Wick gravity framework should be viewed not merely as a theoretical model capable of explaining QPO formation but also as a promising observational channel through which the effects of higher-derivative quantum gravity could be tested in future astrophysical observations.

\section{Conclusions} 

In the weak Lee-Wick region, defined by $S_{1}\!\ge\!2$ and $S_{2}\!\le\!1.5$, the exponential suppression term in the metric function $f(r)$ dominates over the oscillatory correction term. As a consequence, the numerical simulations demonstrate that the shock cone morphology produced in these models does not differ significantly from that of the Schwarzschild spacetime. The curvature of spacetime remains regular and exhibits no substantial deviation from general relativity, leading to a smooth and well-defined accretion structure. As in the standard Schwarzschild BH, the BHL mechanism produces a symmetric shock cone in the downstream region, which maintains a quasi-steady configuration over time while undergoing mild oscillations. These oscillations originate from small perturbations within the post-shock region and appear as coherent harmonic variations in the mass-inflow rate. Furthermore, the opening angle of the shock cone is found to be in close agreement with the results obtained for the Schwarzschild case. This confirms that the Lee-Wick corrections in this regime are weak, the influence of the higher-derivative terms is minimal, and the overall accretion dynamics and the formation of the shock cone are entirely governed by standard general relativistic gravity.

A detailed analysis of test particle motion around the non-rotating Lee-Wick BH shows that parameters $S_1$ and $S_2$ are fundamental determining of the spacetime orbital structure. The effective potential shows clear distinctions between stable and unstable circular orbits and provides a consistent criterion for orbital stability through its minima and maxima. The profiles of energy and angular momentum show that for nearly all parameter regimes $S_1$ and $S_2$ can be increased and energy needed for circular motion is lowered. The same modest impact (decrease) is seen for angular momentum particularly in certain parameter regimes. These tendencies are seen in the radii of the ISCOs, which for large $S_1$ and $S_2$ values show the same trend of decreasing close to a saturation value. In all cases the Lee-Wick modifications exhibit clear and meaningful fundamental changes to the stability and energy criterion and radius of characteristic orbits. Overall, the underlying quantum inspired corrections to geodesic motion in this BH spacetime is underscored by the clear, fundamental and physical changes to the geodesic motion.

In the strong Lee-Wick region, where $S_{1}$ is small and $S_{2}$ is large, the exponential and oscillatory components of the metric function contribute with comparable strength, leading to pronounced modifications in the curvature structure of spacetime. As revealed by the numerical simulations, these strong gravitational corrections produce rapid oscillations in curvature and abrupt variations in the metric gradient near the BH horizon. Such variations strongly influence the dynamics of the accreting matter, causing the downstream shock cone to widen, become asymmetric, and exhibit an increasing level of instability over time. 
Eventually, the initially well-defined shock cone evolves into a bow shock-like configuration that fully surrounds the BH, signifying the transition to a qualitatively different physical regime dominated by oscillatory spacetime curvature. This behavior clearly demonstrates that the results in the strong Lee-Wick domain deviate substantially from the Schwarzschild case, as the accretion flow becomes turbulent and the governing dynamics are dominated by nonlinear gravitational couplings. In particular, the time-dependent evolution of the mass accretion rate in this region shows highly irregular oscillations occurring on very short timescales. The PSD analyzes reveal that these intense instabilities generate multiple frequency components in the range of $10$-$800\,\mathrm{Hz}$, producing a spectrum of QPOs. These signatures confirm that strong mode coupling arises due to the oscillatory curvature term governed by $S_{2}$, resulting in QPOs whose amplitudes and frequency distributions depend sensitively on the Lee-Wick parameters. Therefore, this strong-correction regime provides an important theoretical framework for probing the influence of quantum-gravity effects on the spacetime geometry around BHs, offering a promising avenue for testing the astrophysical implications of Lee-Wick gravity.

The numerical results have shown that the astrophysical manifestations of Lee-Wick gravity vary strongly depending on the parameters $S_{1}$ and $S_{2}$. In the weak correction limit, the morphology of the shock cone closely resembles that of the Schwarzschild solution. 
In this case, the mass accretion flow exhibits smooth behavior with mild temporal modulations capable of explaining the LFQPOs. The numerical findings are in good agreement with the QPOs observed in $X-$ray binary systems such as XTE~J1550-564 and GRS~1915+105. The presence of such LFQPOs indicates that small perturbations within the post-shock region are coherently sustained as harmonic oscillations of the accretion flow, whereas the influence of higher-derivative corrections remains minimal. Therefore, the weak field limit can account for observed astrophysical systems that exhibit stable, low-amplitude QPOs. Furthermore, the stability and regularity of the shock cone in this regime suggest that the observed frequency distribution and amplitude of LFQPOs may impose constraints on the Lee-Wick parameters, potentially setting limits on $S_{1}$ and $S_{2}$ for such sources. However, the numerical results obtained for the strong correction regime reveal distinct observational consequences. The interaction between the spacetime parameters in this domain, characterized by nonlinear gravitational couplings and strong curvature oscillations near the BH horizon, can explain the HFQPOs observed in various sources. In this region, the intense and rapidly varying oscillations provide the physical conditions necessary for HFQPO formation. The PSD analyses of the accretion rate demonstrate that the QPO frequencies extend from a few tens to several hundred hertz. 
Numerically, these HFQPOs arise from the coupling between radial and azimuthal modes, producing frequency ratios such as $3:2$, $2:1$, and $5:2$. These ratios are in excellent agreement with those observed in GRO~J1655-40 (300/450\,Hz) and XTE~J1550-564 (184/276\,Hz). Hence, the oscillatory curvature term governed by $S_{2}$ drives the emergence of nonlinear resonant modes, leading to the transition from LFQPOs to HFQPOs. Consequently, while the weak Lee-Wick regime reproduces the classical LFQPO phenomenology, the strong correction domain naturally explains the appearance of HFQPOs and broadband oscillatory structures. Thus, within a single theoretical framework, Lee-Wick gravity can unify both LFQPO  and HFQPO behaviors, offering a valuable opportunity to investigate quantum gravity induced spacetime modifications in the vicinity of BHs.

\section*{Acknowledgments} 
 All numerical simulations were performed using the Phoenix High Performance Computing facility at the American University of the Middle East (AUM), Kuwait. 

\section*{Data Availability Statement}
The data sets generated and analyzed during the current study were produced using high-performance computing resources. These data are not publicly available due to their large size and computational nature, but are available from the corresponding author upon reasonable request.

\section*{Appendix}
\begin{widetext}
\begin{align*}
\Omega_{r}^{2} &=\frac{e^{-2 r S_1}}{4 r^4 S_1^2 S_2^2} \Big[ 
 - r^4 S_1^8 - 5 r^3 S_1^7 - 9 r^2 S_1^6 - 6 r^4 S_2^2 S_1^6 
 - 4 e^{r S_1} r^3 S_2 \sin(r S_2) S_1^6 - 5 r^3 S_2 \sin(2 r S_2) S_1^6 \\
& + 2 e^{r S_1} r^4 \sin(r S_2) S_2 S_1^6 - 19 r^3 S_2^2 S_1^5 - 6 r S_1^5
 - 4 e^{r S_1} r^3 S_2 \sin(r S_2) S_1^5 - 18 r^2 S_2 \sin(2 r S_2) S_1^5 \\
& + 24 e^{r S_1} r^2 \sin(r S_2) S_2 S_1^5 - 12 r^4 S_2^4 S_1^4
 + 4 e^{r S_1} r^4 \sin(r S_2) S_2^3 S_1^4 - 15 r^2 S_2^2 S_1^4 \\
& - 8 e^{r S_1} r^3 S_2^3 \sin(r S_2) S_1^4 
 - 2 e^{r S_1} r^2 S_2 \sin(r S_2) S_1^4 
 - 15 r^3 S_2^3 \sin(2 r S_2) S_1^4 - 18 r S_2 \sin(2 r S_2) S_1^4 \\
& + 24 e^{r S_1} r \sin(r S_2) S_2 S_1^4 - 3 S_1^4 - 23 r^3 S_2^4 S_1^3
 + 48 e^{r S_1} r^2 \sin(r S_2) S_2^3 S_1^3 - 12 r S_2^2 S_1^3 \\
& - 8 e^{r S_1} r^3 S_2^3 \sin(r S_2) S_1^3
 - 2 e^{r S_1} r S_2 \sin(r S_2) S_1^3 - 36 r^2 S_2^3 \sin(2 r S_2) S_1^3 \\
& - 12 S_2 \sin(2 r S_2) S_1^3 + 24 e^{r S_1} \sin(r S_2) S_2 S_1^3
 - 10 r^4 S_2^6 S_1^2 + 2 e^{r S_1} r^4 \sin(r S_2) S_2^5 S_1^2 \\
& - 3 r^2 S_2^4 S_1^2 + 24 e^{r S_1} r \sin(r S_2) S_2^3 S_1^2
 - 24 e^{2 r S_1} S_2^2 S_1^2 + 4 e^{2 r S_1} r S_2^2 S_1^2 - 6 S_2^2 S_1^2 \\
& - 4 e^{r S_1} r^3 S_2^5 \sin(r S_2) S_1^2
 - 2 e^{r S_1} r^2 S_2^3 \sin(r S_2) S_1^2
 - 15 r^3 S_2^5 \sin(2 r S_2) S_1^2 \\
& - 12 r S_2^3 \sin(2 r S_2) S_1^2
 - 9 r^3 S_2^6 S_1 + 24 e^{r S_1} r^2 \sin(r S_2) S_2^5 S_1
 - 6 r S_2^4 S_1 \\
& + 2 e^{r S_1} r \sin(r S_2) S_2^3 S_1 + 12 \sin(2 r S_2) S_2^3 S_1
 - 2 e^{r S_1} S_2^2 \Big((r-2) r^3 S_1^4 + r (2 S_2^2 r^3 - 4 S_2^2 r^2 + r - 12) S_1^2 \\
& \quad + 2 (r-12) S_1 + r S_2^2 (S_2^2 r^3 - 2 S_2^2 r^2 + r - 12) \Big)
 \cos(r S_2) S_1 - 4 e^{r S_1} r^3 S_2^5 \sin(r S_2) S_1 \\
& - 24 e^{r S_1} S_2^3 \sin(r S_2) S_1 - 18 r^2 S_2^5 \sin(2 r S_2) S_1
 - 3 r^4 S_2^8 + 3 r^2 S_2^6 + 6 r \sin(2 r S_2) S_2^5 \\
& - 3 S_2^4 - 5 r^3 S_2^7 \sin(2 r S_2)
 + \cos(2 r S_2) \Big(r^4 S_1^8 + 5 r^3 S_1^7 + r^2 (4 r^2 S_2^2 + 9) S_1^6 \\
& + 3 r (5 r^2 S_2^2 + 2) S_1^5 + (6 r^4 S_2^4 + 9 r^2 S_2^2 + 3) S_1^4
 + 3 r S_2^2 (5 r^2 S_2^2 - 4) S_1^3 \\
& + S_2^2 (4 r^4 S_2^4 - 9 r^2 S_2^2 - 18) S_1^2
 + r S_2^4 (5 r^2 S_2^2 - 18) S_1
 + S_2^4 (r^4 S_2^4 - 9 r^2 S_2^2 + 3) \Big)
\Big],\\
\Omega_{\theta}^{2} &= \frac{e^{-r S_1}}{2 r^3 S_1 S_2} (-\left(r^2 S_1^4+2 r^2 S_2^2 S_1^2+r^2 S_2^4+r S_1^3+r S_2^2 S_1+S_1^2-S_2^2\right) \sin \left(r S_2\right)+2 S_1 S_2 e^{r S_1} \\ &- \left(S_2 \left(r S_1^2+r S_2^2+2 S_1\right) \cos \left(r S_2\right)\right)),\\
\Omega_\phi^2 &= \frac{e^{-r S_1}}{2 r^3 S_1 S_2} (-\left(r^2 S_1^4+2 r^2 S_2^2 S_1^2+r^2 S_2^4+r S_1^3+r S_2^2 S_1+S_1^2-S_2^2\right) \sin \left(r S_2\right)+2 S_1 S_2 e^{r S_1} \\ &- \left(S_2 \left(r S_1^2+r S_2^2+2 S_1\right) \cos \left(r S_2\right)\right)).
\end{align*}
\end{widetext}

\bibliographystyle{apsrev4-1}  

\end{document}